\documentclass[12pt,letterpaper]{article}

\usepackage{iftex}
\ifPDFTeX
  \usepackage[T1]{fontenc}
  \usepackage{newtxtext,newtxmath}
\else
  \usepackage{fontspec}
  \usepackage{newtxmath}
\fi
\usepackage[top=1in,bottom=1in,left=0.75in,right=0.75in,footskip=20pt]{geometry}
\usepackage{setspace}
\usepackage[super,sort&compress]{natbib}
\AtBeginDocument{\setcitestyle{citesep={,}}}
\usepackage{microtype}
\usepackage{lineno}
\modulolinenumbers[1]
\usepackage{amsmath}
\usepackage{graphicx}
\usepackage{float}
\usepackage{booktabs}
\usepackage{tabularx}
\usepackage{array}
\usepackage{multirow}
\usepackage{threeparttable}
\usepackage{longtable}
\usepackage{caption}
\usepackage{titlesec}
\usepackage{hyperref}
\usepackage{makecell}
\hypersetup{hidelinks}
\titleformat{\section}{\fontsize{14}{18}\selectfont\bfseries}{}{0pt}{}
\titleformat{\subsection}{\normalsize\bfseries}{}{0pt}{}
\titleformat{\subsubsection}{\normalsize\bfseries}{}{0pt}{}
\titlespacing*{\section}{0pt}{28pt}{4pt}
\titlespacing*{\subsection}{0pt}{17pt}{5pt}
\titlespacing*{\subsubsection}{0pt}{14pt}{4pt}

\newcolumntype{Y}{>{\raggedright\arraybackslash}X}

\newcommand{\qfdr}{q_{\mathrm{FDR}}}

\usepackage{xcolor}

\newcolumntype{P}[1]{>{\raggedright\arraybackslash}p{#1}}
\newcolumntype{C}[1]{>{\centering\arraybackslash}p{#1}}

\begin{document}

\thispagestyle{plain}
\vspace*{-0.45in}
\begin{center}
{\fontsize{17}{22}\selectfont\bfseries
Task-evoked eye movements as candidate digital biomarkers of depression severity in adolescents and young adults
\par}
\end{center}

\vspace{0.15in}
% Author block for UMLPA-GCN npj-style submission.
\begin{center}
Youxiang Sun$^{1,2,\dagger}$,
Yujie Nie$^{1,2,\dagger}$,
Zhuoru Wu$^{4,\dagger}$,
Mengyuan Yu$^{3}$,
Wenkong Wang$^{1,2}$,
Xiang Li$^{1,2}$,
Yumei Wang$^{3,*}$,
Xin Ma$^{1,2,*}$,
and Lizhou Fan$^{4,*}$
\par
\end{center}

\vspace{0.12in}

{\fontsize{9.7}{10.7}\selectfont
\setlength{\parindent}{0pt}
\setlength{\parskip}{0pt}
$^{1}$School of Control Science and Engineering, Shandong University, Jinan, Shandong, China\par
$^{2}$Engineering Research Center of Intelligent Unmanned System, Ministry of Education, China\par
$^{3}$Department of Psychology, Shandong Provincial Hospital Affiliated to Shandong First Medical University, Jinan, Shandong, China\par
$^{4}$Department of Psychiatry, The Chinese University of Hong Kong, Shatin, Hong Kong SAR, China\par

\vspace{0.10in}
$^{\dagger}$These authors contributed equally: Youxiang Sun, Yujie Nie, and Zhuoru Wu.\par
$^{*}$Correspondence should be addressed to Yumei Wang (wangym@bjmu.edu.cn), Xin Ma (maxin@sdu.edu.cn), or Lizhou Fan (leofan@cuhk.edu.hk)
}

\clearpage
% \linenumbers
\nolinenumbers % 关闭行号
\noindent{\fontsize{14}{18}\selectfont\bfseries Abstract:}\par
\vspace{8pt}

Objective digital measures of depression severity could complement conventional symptom assessments and facilitate more timely clinical evaluation. Here, we investigated whether task-evoked eye movements provide interpretable behavioral markers of depression severity in adolescents and young adults. In this single-center, cross-sectional study, 155 participants aged 14--20 years completed emotional-spatial free-viewing and memory-guided recall tasks and were stratified by 17-item Hamilton Depression Rating Scale scores. We examined 19 prespecified eye-movement features across six task-informed domains using covariate-adjusted ordinal-trend models. Greater depression severity was associated with lower gaze-allocation entropy (standardized $\beta=-0.599$), fewer inter-area switches ($\beta=-0.526$), longer first target-entry latency ($\beta=0.572$), and lower target dwell proportion ($\beta=-0.678$; all false-discovery-rate-adjusted $q<0.001$). Emotion-specific attentional-bias measures were not associated with severity, whereas associations involving fixation dispersion weakened after adjustment for anxiety symptoms. In a complementary machine-learning analysis incorporating 44 candidate eye-movement features and demographic covariates, a three-stage cascade classifier achieved the highest four-class stratification performance (accuracy, 0.665; macro-F1, 0.670; area under the receiver operating characteristic curve, 0.869), outperforming single-stage and two-stage approaches. Together, these findings identify altered visual exploration and memory-guided gaze control as candidate digital biomarkers of depression severity, rather than generalized negative attentional bias. These results support further prospective evaluation of task-evoked eye tracking as a complementary approach for timely severity assessment and clinical decision support.

\vspace{0.45cm}
\noindent\textbf{Keywords:} adolescent depression, eye tracking, digital biomarker, visual attention, spatial memory, psychomotor behavior

\newpage

\section{Introduction}

Depression during adolescence is common, disabling, and associated with recurrent illness, impaired social and educational functioning, self-harm, and suicide risk \cite{Shorey2022,Friedrich2017,Clayborne2019,Johnson2018,Grossberg2023,Dwyer2020}. This developmental period also involves substantial maturation of emotion regulation, executive control, memory, and social cognition \cite{Fitzpatrick2023,Rice2019}. Clinical assessment therefore needs to capture not only affective symptoms but also the cognitive and behavioral changes through which depression may interfere with daily functioning.

Depression has been linked to alterations in cognitive control, attention, memory, and processing of socially salient emotional information \cite{Beck2008,LeMoult2019,Everaert2017}. Eye tracking provides temporally resolved measurement of overt visual attention and oculomotor behavior and can distinguish initial orienting, sustained allocation, scanpath organization, and fixation-related processing \cite{Carter2020,Skaramagkas2023,Hessels2025}. These properties make eye movement a candidate task-based digital behavioral measure, although individual metrics are not specific to a single latent cognitive process and require interpretation within the demands of the task that generated them.

Most depression eye-tracking research has focused on emotional attentional bias. Meta-analytic evidence suggests that clinical depression is more consistently associated with prolonged maintenance on dysphoric material and reduced attention to positive material than with altered initial orienting \cite{Armstrong2012,Suslow2020}. Studies using emotional faces and other affective stimuli have also reported mood-congruent gaze patterns and associations with particular symptom dimensions \cite{Akram2021,Klawohn2020,Sun2024,Liu2025}. However, findings vary across stimulus sets, viewing durations, and task instructions, indicating that emotional valence alone may not account for depression-related gaze behavior.

Broader alterations have been reported in visual exploration, fixation control, saccadic behavior, and memory-related processing \cite{Carvalho2015,Toh2011,Takahashi2021,Wang2022,Gao2023,Zhang2022}. Eye-tracking studies involving memory tasks further suggest inefficient cue detection, impaired recognition of learned face--scene associations, and reduced allocation to remembered targets in depression \cite{Chen2013,Nemeth2016}. In adolescents, eye-movement characteristics have also been associated with cognitive functioning in major depressive disorder \cite{Lu2024}. Together, these findings motivate paradigms that jointly examine how participants distribute gaze while encoding information and how they use that information to guide subsequent behavior.

Three limitations remain. First, case--control comparisons cannot establish whether a feature changes progressively with symptom burden. Second, single-task studies and isolated metrics provide limited information about whether observed effects arise during visual sampling, scanpath organization, memory-guided orienting, or maintenance after target acquisition. Third, statistical associations are difficult to interpret when eye-movement features are not prespecified according to task demands and plausible behavioral processes \cite{Smith2018,Hessels2025}.

We therefore combined emotional-spatial free-viewing with memory-guided recall and organized 19 primary eye-movement features into six prespecified domains: visual scanning coverage, emotion-related attentional bias, gaze transition dynamics, memory-guided target acquisition, target-area gaze maintenance, and fixation stability. We tested ordered associations across four HAMD-17 symptom-severity strata and evaluated dimensional consistency using continuous HAMD-17 scores. We hypothesized that greater symptom severity would be associated with less comprehensive and less organized visual sampling, greater negative attentional maintenance, delayed target acquisition, weaker target maintenance, and poorer fixation stability.
Beyond these dimensional tests, we used the task-informed feature set as input to an exploratory machine-learning analysis of whether the candidate features jointly separate the four severity strata and whether hierarchical (cascade) decision rules improve on single-stage classification.
The study was designed to characterize task-related behavioral correlates of symptom severity, not to develop or validate a diagnostic test.

\section{Results}

\subsection{Cohort characteristics and data quality}

Of the 203 recruited participants, 48 were excluded because their proportion of valid eye-tracking data was below the prespecified threshold of 80\%, leaving 155 participants for analysis. The analytic cohort comprised 32 participants with HAMD-17 scores of 0--7, 39 with scores of 8--16, 49 with scores of 17--23, and 35 with scores of 24 or higher. Mean age and years of education differed across severity groups, whereas the proportion of male participants did not (Table~\ref{tab:characteristics}). 

The observed severity gradient must be interpreted in relation to the recruitment design. Participants without depression were recruited from the community, whereas participants with major depressive disorder were recruited from a clinical service. Consequently, comparisons that span the no-depression and symptomatic groups may reflect both symptom severity and recruitment-source or diagnostic-status differences. 

\begin{table}[!t]
\centering
\begin{threeparttable}
\caption{\textbf{Demographic and clinical characteristics across HAMD-17 symptom-severity groups.}}
\label{tab:characteristics}
\small
\begin{tabular}{lcccccc}
\toprule
 & No depression & Mild & Moderate & Severe & Statistic & $p$ \\
 & ($n=32$) & ($n=39$) & ($n=49$) & ($n=35$) & & \\
\midrule
Age, years & 17.22 (2.56) & 17.23 (2.24) & 16.06 (2.67) & 15.60 (2.63) & $H=10.84$ & 0.013 \\
Male sex, no. (\%) & 14 (43.8) & 16 (41.0) & 17 (34.7) & 10 (28.6) & $\chi^2=2.08$ & 0.557 \\
Education, years & 11.34 (2.35) & 11.18 (2.04) & 10.14 (2.37) & 9.69 (2.27) & $H=12.94$ & 0.005 \\
HAMD-17 score & 4.28 (2.44) & 12.51 (3.09) & 19.16 (2.22) & 27.06 (3.87) & $H=144.02$ & $<0.001$ \\
HAMA score & 5.09 (5.01) & 9.31 (5.05) & 14.10 (7.23) & 18.37 (8.11) & $H=53.47$ & $<0.001$ \\
\bottomrule
\end{tabular}
\begin{tablenotes}[flushleft]\footnotesize
\item Continuous variables are mean (SD). Group comparisons used Kruskal--Wallis tests for continuous variables and a chi-square test for sex. HAMA, Hamilton Anxiety Rating Scale; HAMD-17, 17-item Hamilton Depression Rating Scale.
\end{tablenotes}
\end{threeparttable}
\end{table}

\subsection{Ordered-severity associations in visual scanning and gaze transition dynamics}

A total of 44 eye-movement features were extracted to assess their associations with ordered HAMD-17 symptom severity. Operational definitions and calculation procedures for all features are provided in Table~\ref{tab:s1_feature_definitions}. Table~\ref{tab:primary} presents the prespecified primary features, whereas results for the remaining supplementary features are reported in Table~\ref{tab:s2_ordered_severity}.

\begin{figure}[!t]
\centering
\includegraphics[width=\textwidth]{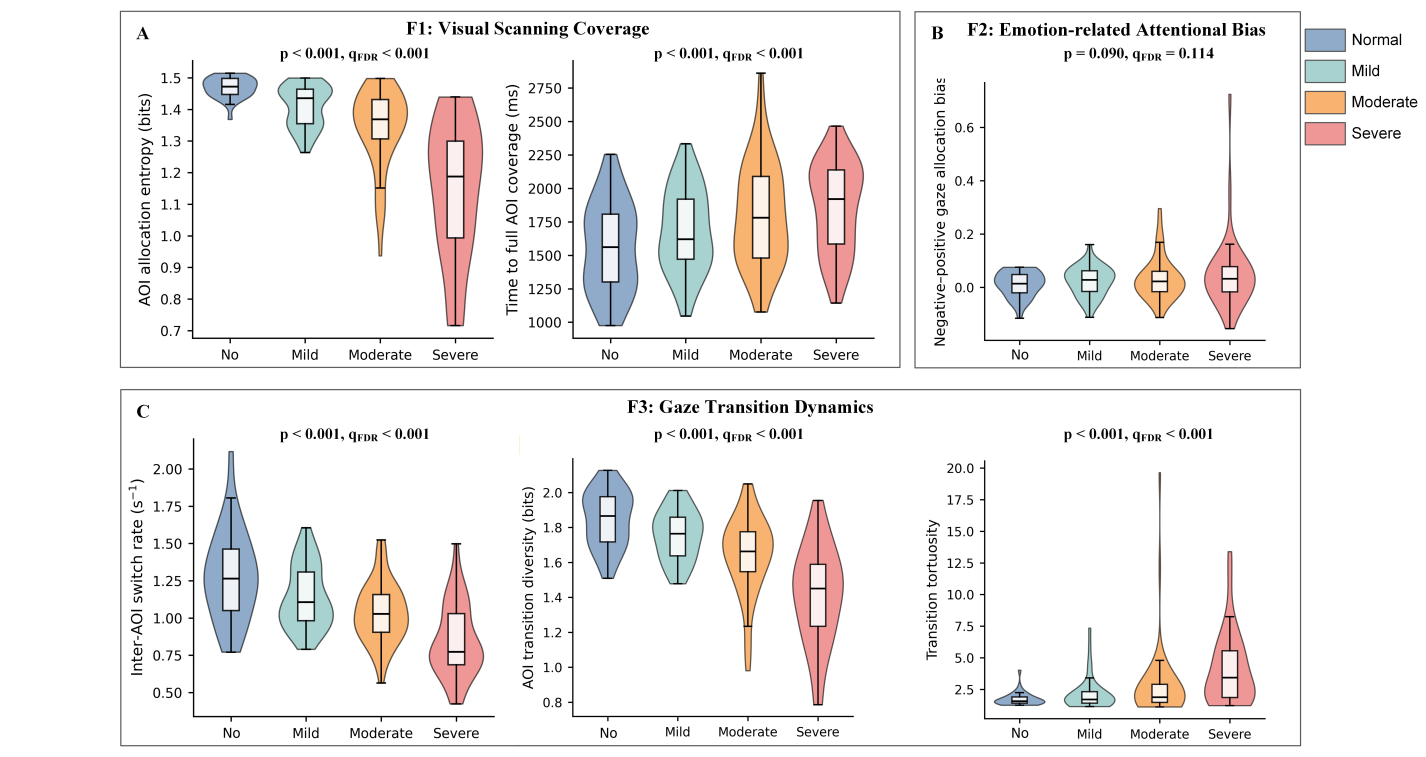}
\caption{\textbf{Representative free-viewing eye-movement features across HAMD-17 symptom-severity groups.} Violin plots with embedded boxplots show unadjusted participant-level distributions for selected features from visual scanning coverage, emotion-related attentional bias, and gaze transition dynamics. Displayed $p$ values were obtained from ordered-severity regression models adjusted for age, sex, and education, with HC3 standard errors. The displayed $q$ values were adjusted across the 19 prespecified primary features. }
\label{fig:freeview}
\end{figure}

Severity-related associations during free-viewing were concentrated in visual scanning coverage and gaze transition dynamics (Table~\ref{tab:primary}). All three primary visual-scanning features were associated with the ordered severity variable after false-discovery-rate correction. Each one-level increase in severity was associated with a 0.305-SD longer time to full area-of-interest (AOI) coverage (95\% CI, 0.165--0.445), a 0.447-SD lower first-pass AOI coverage ratio (95\% CI, $-0.604$ to $-0.291$), and a 0.599-SD lower AOI allocation entropy (95\% CI, $-0.733$ to $-0.466$; all $\qfdr<0.001$). These estimates indicate slower, less complete, and less balanced sampling of the three emotional face locations as symptom severity increased (Fig.~\ref{fig:freeview}).

Gaze transitions showed a similar ordered pattern. Greater severity was associated with a lower inter-AOI switch rate ($\beta=-0.526$, 95\% CI, $-0.674$ to $-0.379$), lower transition diversity ($\beta=-0.567$, 95\% CI, $-0.705$ to $-0.429$), and greater transition tortuosity ($\beta=0.334$, 95\% CI, 0.186--0.482; all $\qfdr<0.001$). Within-AOI gaze dispersion was not significant after correction. The convergence of lower switch frequency, lower transition diversity, and greater tortuosity supports an association with less efficient scanpath organization rather than a simple reduction in total looking.

\begin{table}[!t]
\centering
\begin{threeparttable}
\caption{\textbf{Associations between ordered HAMD-17 symptom severity and prespecified eye-movement features.}}
\label{tab:primary}
\scriptsize
\begin{tabular}{llrrr}
\toprule
Feature & Domain & Standardized $\beta$ (95\% CI) & $p$ & $\qfdr$ \\
\midrule
Time to full AOI coverage & F1 & 0.305 (0.165, 0.445) & $<0.001$ & $<0.001$ \\
First-pass AOI coverage ratio & F1 & $-0.447$ ($-0.604$, $-0.291$) & $<0.001$ & $<0.001$ \\
AOI allocation entropy & F1 & $-0.599$ ($-0.733$, $-0.466$) & $<0.001$ & $<0.001$ \\
Negative--positive gaze allocation bias & F2 & 0.124 ($-0.019$, 0.266) & 0.090 & 0.114 \\
Negative AOI visit duration & F2 & $-0.065$ ($-0.238$, 0.109) & 0.466 & 0.492 \\
Negative stickiness after first hit & F2 & $-0.020$ ($-0.184$, 0.145) & 0.813 & 0.813 \\
Inter-AOI switch rate & F3 & $-0.526$ ($-0.674$, $-0.379$) & $<0.001$ & $<0.001$ \\
AOI transition diversity & F3 & $-0.567$ ($-0.705$, $-0.429$) & $<0.001$ & $<0.001$ \\
Transition tortuosity & F3 & 0.334 (0.186, 0.482) & $<0.001$ & $<0.001$ \\
Within-AOI gaze dispersion & F3 & 0.131 ($-0.013$, 0.275) & 0.075 & 0.102 \\
First target-entry latency & R1 & 0.572 (0.442, 0.702) & $<0.001$ & $<0.001$ \\
Initial target-entry accuracy & R1 & $-0.600$ ($-0.706$, $-0.493$) & $<0.001$ & $<0.001$ \\
First target-fixation latency & R1 & 0.580 (0.454, 0.706) & $<0.001$ & $<0.001$ \\
Target dwell proportion & R2 & $-0.678$ ($-0.778$, $-0.578$) & $<0.001$ & $<0.001$ \\
Target lock-on ratio & R2 & $-0.437$ ($-0.575$, $-0.299$) & $<0.001$ & $<0.001$ \\
Post-hit off-target ratio & R2 & 0.224 (0.103, 0.344) & $<0.001$ & $<0.001$ \\
Target-area fixation dispersion & R3 & 0.179 (0.022, 0.336) & 0.025 & 0.037 \\
Within-target fixation step size & R3 & 0.097 ($-0.058$, 0.251) & 0.219 & 0.245 \\
Target-area stability ellipse area & R3 & 0.125 ($-0.047$, 0.297) & 0.153 & 0.182 \\
\bottomrule
\end{tabular}
\begin{tablenotes}[flushleft]\footnotesize
\item Each feature was standardized before fitting a separate linear regression model. Ordered severity was coded 0--3 and models adjusted for age, sex, and education. HC3 standard errors were used. The coefficient is the SD difference in the eye-movement feature associated with a one-level increase in severity. FDR correction was applied across all 19 primary features. AOI, area of interest; CI, confidence interval; F1, visual scanning coverage; F2, emotion-related attentional bias; F3, gaze transition dynamics; R1, memory-guided target acquisition; R2, target-area gaze maintenance; R3, fixation stability.
\end{tablenotes}
\end{threeparttable}
\end{table}

\subsection{Continuous HAMD-17 associations and sensitivity analyses}

Analyses using continuous HAMD-17 scores showed generally similar directions to the ordered-severity results for the F1, F3, R1, and R2 domains, whereas associations in the F2 domain remained limited. The primary-feature results are summarized in Table~\ref{tab:continuous_primary}, whereas complete results for supplementary features are provided in Table~\ref{tab:s3_continuous_hamd17}. 

In the F1 domain, higher continuous HAMD-17 scores were associated with longer time to full AOI coverage ($\beta=0.299$, 95\% CI, 0.157--0.441), lower first-pass AOI coverage ratio ($\beta=-0.435$, 95\% CI, $-0.586$ to $-0.283$), and lower AOI allocation entropy ($\beta=-0.549$, 95\% CI, $-0.682$ to $-0.416$; all $\qfdr<0.001$). Spearman correlations showed the same directions ($\rho_s=0.278$, $-0.405$, and $-0.651$, respectively; all $\qfdr<0.001$).

In the F3 domain, higher HAMD-17 scores were associated with a lower inter-AOI switch rate ($\beta=-0.500$, 95\% CI, $-0.652$ to $-0.347$), lower AOI transition diversity ($\beta=-0.506$, 95\% CI, $-0.644$ to $-0.369$), and greater transition tortuosity ($\beta=0.270$, 95\% CI, 0.128--0.412; all $\qfdr<0.001$). The corresponding Spearman correlations showed the same directions ($\rho_s=-0.465$, $-0.516$, and 0.326, respectively; all $\qfdr<0.001$). Within-AOI gaze dispersion was not significant in either analysis.

In the R1 domain, higher HAMD-17 scores were associated with longer first target-entry latency ($\beta=0.515$, 95\% CI, 0.351--0.679), lower initial target-entry accuracy ($\beta=-0.562$, 95\% CI, $-0.682$ to $-0.441$), and longer first target-fixation latency ($\beta=0.515$, 95\% CI, 0.350--0.681; all $\qfdr<0.001$). In the R2 domain, higher HAMD-17 scores were associated with lower target dwell proportion ($\beta=-0.624$, 95\% CI, $-0.750$ to $-0.498$), lower target lock-on ratio ($\beta=-0.369$, 95\% CI, $-0.539$ to $-0.199$), and higher post-hit off-target ratio ($\beta=0.199$, 95\% CI, 0.076--0.323; $\qfdr=0.003$ for post-hit off-target ratio and $<0.001$ for the other two). Spearman correlations for these six R1 and R2 features were also significant and showed the same directions.

Associations in the F2 domain were weaker. Negative--positive gaze allocation bias was not significant in the Spearman analysis ($\rho_s=0.114$, $\qfdr=0.206$), although it reached the FDR-corrected threshold in the adjusted regression model ($\beta=0.152$, 95\% CI, 0.011--0.293; $\qfdr=0.049$). Negative AOI visit duration and negative stickiness after first hit were not significant in either analysis. In the R3 domain, target-area fixation dispersion was positively associated with continuous HAMD-17 scores ($\rho_s=0.225$, $\qfdr=0.008$; $\beta=0.221$, 95\% CI, 0.069--0.374; $\qfdr=0.007$), whereas within-target fixation step size and target-area stability ellipse area were not significant after correction. 

\begin{table*}[!t]
\centering
\begin{threeparttable}
\caption{\textbf{Associations between continuous HAMD-17 scores and prespecified eye-movement features.}}
\label{tab:continuous_primary}
\scriptsize

\begin{tabular*}{\textwidth}{@{\extracolsep{\fill}}llrrrrrr@{}}
\toprule
& & \multicolumn{3}{c}{Spearman correlation}
& \multicolumn{3}{c}{Adjusted regression model} \\
\cmidrule(lr){3-5}
\cmidrule(lr){6-8}
Feature & Domain & $\rho_s$ & $p$ & $q_{\mathrm{FDR}}$
& Standardized $\beta$ (95\% CI) & $p$ & $q_{\mathrm{FDR}}$ \\
\midrule

Time to full AOI coverage
& F1 & 0.278 & $<0.001$ & $<0.001$
& 0.299 (0.157, 0.441) & $<0.001$ & $<0.001$ \\

First-pass AOI coverage ratio
& F1 & $-0.405$ & $<0.001$ & $<0.001$
& $-0.435$ ($-0.586$, $-0.283$) & $<0.001$ & $<0.001$ \\

AOI allocation entropy
& F1 & $-0.651$ & $<0.001$ & $<0.001$
& $-0.549$ ($-0.682$, $-0.416$) & $<0.001$ & $<0.001$ \\

Negative--positive gaze allocation bias
& F2 & 0.114 & 0.158 & 0.206
& 0.152 (0.011, 0.293) & 0.035 & 0.049 \\

Negative AOI visit duration
& F2 & $-0.095$ & 0.242 & 0.273
& 0.033 ($-0.163$, 0.230) & 0.740 & 0.776 \\

Negative stickiness after first hit
& F2 & 0.021 & 0.798 & 0.817
& 0.045 ($-0.122$, 0.212) & 0.598 & 0.675 \\

Inter-AOI switch rate
& F3 & $-0.465$ & $<0.001$ & $<0.001$
& $-0.500$ ($-0.652$, $-0.347$) & $<0.001$ & $<0.001$ \\

AOI transition diversity
& F3 & $-0.516$ & $<0.001$ & $<0.001$
& $-0.506$ ($-0.644$, $-0.369$) & $<0.001$ & $<0.001$ \\

Transition tortuosity
& F3 & 0.326 & $<0.001$ & $<0.001$
& 0.270 (0.128, 0.412) & $<0.001$ & $<0.001$ \\

Within-AOI gaze dispersion
& F3 & 0.103 & 0.202 & 0.233
& 0.054 ($-0.099$, 0.208) & 0.489 & 0.566 \\

First target-entry latency
& R1 & 0.586 & $<0.001$ & $<0.001$
& 0.515 (0.351, 0.679) & $<0.001$ & $<0.001$ \\

Initial target-entry accuracy
& R1 & $-0.597$ & $<0.001$ & $<0.001$
& $-0.562$ ($-0.682$, $-0.441$) & $<0.001$ & $<0.001$ \\

First target-fixation latency
& R1 & 0.597 & $<0.001$ & $<0.001$
& 0.515 (0.350, 0.681) & $<0.001$ & $<0.001$ \\

Target dwell proportion
& R2 & $-0.666$ & $<0.001$ & $<0.001$
& $-0.624$ ($-0.750$, $-0.498$) & $<0.001$ & $<0.001$ \\

Target lock-on ratio
& R2 & $-0.482$ & $<0.001$ & $<0.001$
& $-0.369$ ($-0.539$, $-0.199$) & $<0.001$ & $<0.001$ \\

Post-hit off-target ratio
& R2 & 0.290 & $<0.001$ & $<0.001$
& 0.199 (0.076, 0.323) & 0.002 & 0.003 \\

Target-area fixation dispersion
& R3 & 0.225 & 0.005 & 0.008
& 0.221 (0.069, 0.374) & 0.004 & 0.007 \\

Within-target fixation step size
& R3 & 0.114 & 0.159 & 0.206
& 0.131 ($-0.028$, 0.291) & 0.106 & 0.141 \\

Target-area stability ellipse area
& R3 & 0.135 & 0.096 & 0.133
& 0.164 ($-0.022$, 0.351) & 0.084 & 0.116 \\

\bottomrule
\end{tabular*}

\begin{tablenotes}[flushleft]
\footnotesize
\item $\rho_s$ denotes the Spearman rank correlation coefficient between continuous HAMD-17 total scores and eye-movement features. Adjusted $\beta$ values were estimated using separate linear regression models with standardized HAMD-17 total score as the predictor and each standardized eye-movement feature as the outcome, adjusting for age, sex, and years of education. Robust standard errors were used. The $\beta$ coefficient represents the SD difference in the eye-movement feature associated with a one-SD increase in HAMD-17 total score. FDR correction was applied separately to the Spearman correlation tests and the adjusted regression tests across all 44 eye-movement features, including 19 primary and 25 supplementary features. AOI, area of interest; CI, confidence interval; FDR, false discovery rate; F1, visual scanning coverage; F2, emotion-related attentional bias; F3, gaze transition dynamics; R1, memory-guided target acquisition; R2, target-area gaze maintenance; R3, fixation stability.
\end{tablenotes}

\end{threeparttable}
\end{table*}

Additional adjustment for HAMA scores in Table~\ref{tab:s4_sensitivity} generally preserved the F1, F3, R1, and R2 associations, with most estimates remaining similar in direction and magnitude. In contrast, the association for target-area fixation dispersion was attenuated and was no longer significant after FDR correction. Excluding feature-specific extreme observations ($|z|>3$) yielded a similar overall pattern, with the principal F1, F3, R1, and R2 associations remaining significant.

\subsection{Emotion-related attentional bias across symptom severity}

None of the three primary emotion-related attentional-bias features remained significant after correction. The negative--positive gaze allocation bias had $\beta=0.124$ (95\% CI, $-0.019$ to 0.266; $\qfdr=0.114$), negative-AOI visit duration had $\beta=-0.065$ (95\% CI, $-0.238$ to 0.109; $\qfdr=0.492$), and negative stickiness after first hit had $\beta=-0.020$ (95\% CI, $-0.184$ to 0.145; $\qfdr=0.813$). Thus, the broad severity gradients in coverage and transition organization were not accompanied by evidence of a generalized shift toward negative faces in this memory-relevant viewing context.

\begin{figure}[!t]
\centering
\includegraphics[width=\textwidth]{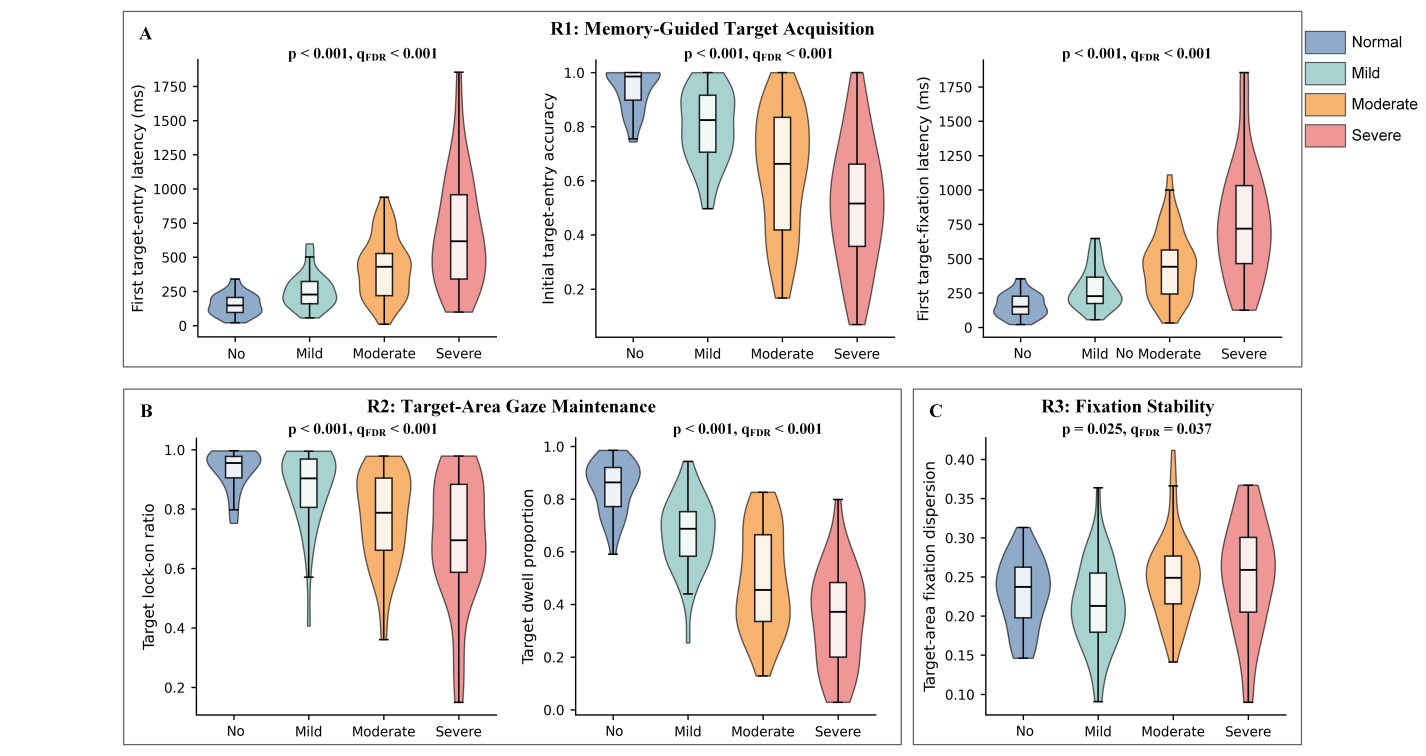}
\caption{\textbf{Representative memory-guided recall features across HAMD-17 symptom-severity groups.} Violin plots with embedded boxplots show unadjusted participant-level distributions for selected features from target acquisition, target-area gaze maintenance, and fixation stability. Displayed $p$ values were obtained from ordered-severity regression models adjusted for age, sex, and education, with HC3 standard errors. The displayed $q$ values were adjusted across the 19 prespecified primary features. }
\label{fig:recall}
\end{figure}

\subsection{Ordered-severity associations in memory-guided target acquisition and gaze maintenance}

Overall recall performance decreased across the four groups ($H=71.97$, $p<0.001$). Group-specific numbers of correct recall trials are provided in Table~\ref{tab:s5_recall_performance}.

All three target-acquisition features were strongly associated with ordered severity (Fig.~\ref{fig:recall}). Each severity step was associated with a 0.572-SD longer first target-entry latency (95\% CI, 0.442--0.702), a 0.600-SD lower initial target-entry accuracy (95\% CI, $-0.706$ to $-0.493$), and a 0.580-SD longer first target-fixation latency (95\% CI, 0.454--0.706; all $\qfdr<0.001$). These findings indicate that participants with greater symptom severity were slower and less accurate when orienting to the remembered target location.

Target maintenance after acquisition showed the largest standardized association. Greater severity was associated with a lower target dwell proportion ($\beta=-0.678$, 95\% CI, $-0.778$ to $-0.578$), lower target lock-on ratio ($\beta=-0.437$, 95\% CI, $-0.575$ to $-0.299$), and higher post-hit off-target ratio ($\beta=0.224$, 95\% CI, 0.103--0.344; all $\qfdr<0.001$). By contrast, evidence for local fixation stability was limited. Target-area fixation dispersion was associated with severity in the primary model ($\beta=0.179$, 95\% CI, 0.022--0.336; $\qfdr=0.037$), whereas within-target fixation step size and stability ellipse area were not.

% \begin{revblock}
\subsection{Machine-learning classification of the four severity strata}
As an exploratory analysis, we asked whether the candidate eye-movement features jointly separate the four severity strata. 
Model input comprised the 44 candidate features (25 free-viewing and 19 recall features, of which the 19 prespecified primary features are a subset) together with age, sex, and education, giving 47 inputs per participant. 
We evaluated 21 classifiers (three baseline models, the TabPFN transformer, fourteen AutoGluon configurations, and three simple-average ensembles; Methods) under four protocols: single-stage four-class classification; a two-stage cascade that first separates no/mild from moderate/severe strata and then splits each branch (2+2); a two-stage cascade that first separates no depression from the symptomatic strata and then classifies mild, moderate, and severe (1+3); and a three-stage cascade of successive binary decisions (2+2+2).

\begin{figure}[htbp]
\centering
\includegraphics[height=0.8\textheight]{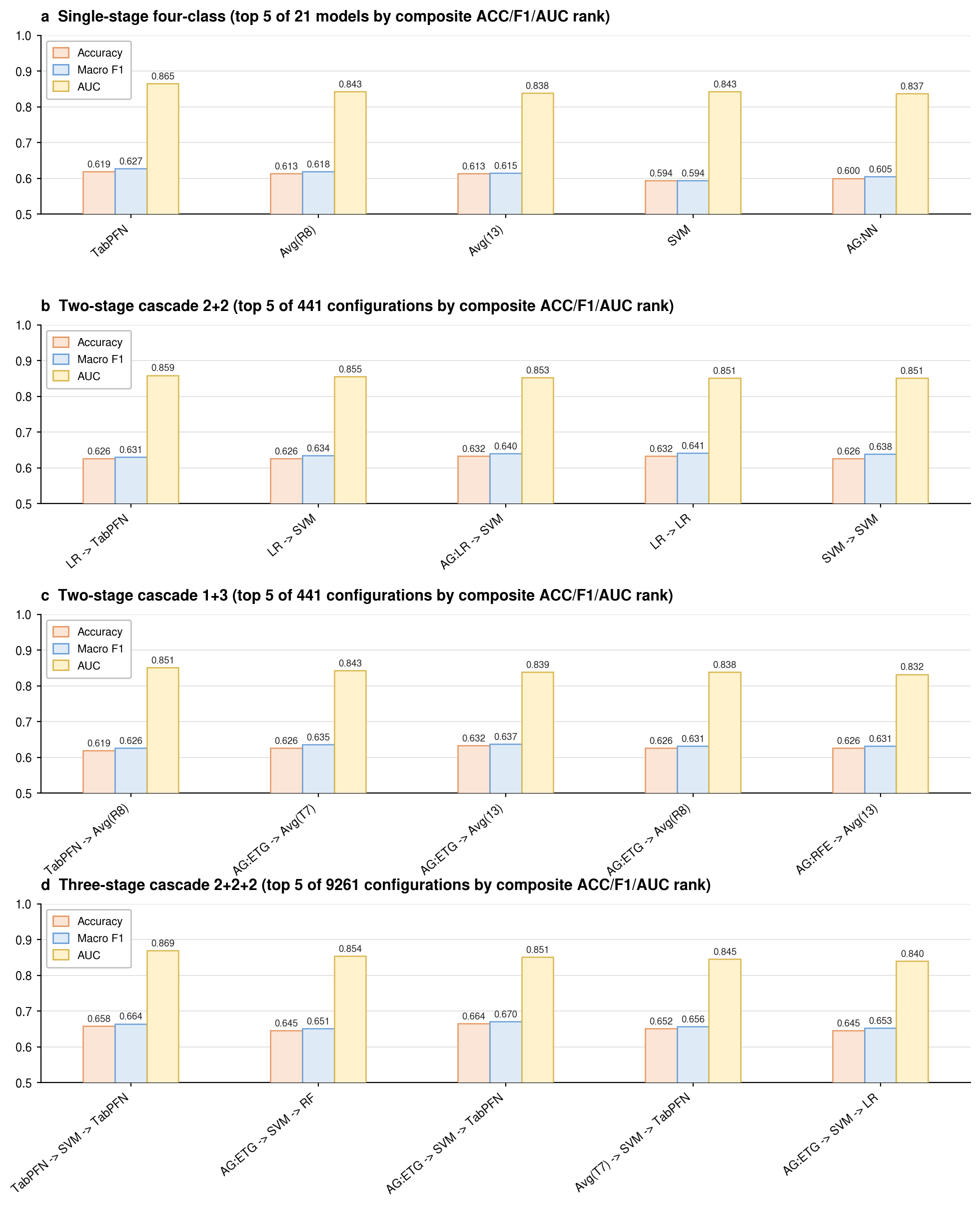}
\caption{\textbf{Top-5 machine-learning configurations for the four classification protocols.} \textbf{a}, single-stage four-class (21 models); \textbf{b}, two-stage cascade 2+2 (441 configurations); \textbf{c}, two-stage cascade 1+3 (441 configurations); \textbf{d}, three-stage cascade 2+2+2 (9261 configurations). For each protocol, configurations are ordered by the sum of their ranks across accuracy, macro F1, and AUC (top 5 shown, left to right); grouped bars show the three metric values of each configuration. Values are means over five folds of stratified cross-validation. AG, AutoGluon; NN, NeuralNetTorch; ETE, ExtraTreesEntr; ETG, ExtraTreesGini; RFE, RandomForestEntr; LR, logistic regression; AG:LR, AutoGluon LinearModel; RF, random forest; SVM, support vector machine (RBF); TabPFN, TabPFN transformer; Avg(13), simple-average ensemble of 13 base AutoGluon models; Avg(T7), ensemble of the seven tree models; Avg(R8), ensemble of one representative per model family.}
\label{fig:ml_settings}
\end{figure}

Performance was the mean over five folds of stratified cross-validation. Configurations were ranked by the sum of their ranks across accuracy, macro F1, and one-vs-rest macro AUC (rank 1 = best per metric), and Table~\ref{tab:ml} lists the top five configurations per protocol (grouped bar plots in Fig.~\ref{fig:ml_settings}); each listed configuration ranks within the top five for at least two of the three metrics. 
The three-stage cascade occupied the top of the composite ranking, with TabPFN to support vector machine (RBF) to TabPFN first (accuracy 0.658, macro F1 0.664, AUC 0.869). The highest individual metric value across all protocols was accuracy 0.665, macro F1 0.670, and AUC 0.851 (AutoGluon ExtraTreesGini to support vector machine (RBF) to TabPFN). Both two-stage cascades and single-stage classification were lower (best accuracy 0.632, 0.632, and 0.619, respectively). In an exploratory two-class task that asked only whether participants fell in the no/mild or the moderate/severe strata, the best classifier (logistic regression) reached accuracy 0.839 (macro F1 0.835) and AUC 0.929 (Fig.~\ref{fig:ml_binary}). The coarse no/mild versus moderate/severe boundary is therefore well supported by these features, whereas fine-grained separation, in particular between the moderate and severe strata, remains difficult. Because best-of-many configurations were selected from 441 or 9261 candidates, these values are an upper-bound characterization of what the feature set can support rather than the performance of a single prespecified model.
% \end{revblock}

\begin{table}[!t]
% \begin{revblock}
\centering
\begin{threeparttable}
\caption{\textbf{Top-5 machine-learning configurations per protocol by composite ACC/F1/AUC rank.}}
\label{tab:ml}
\footnotesize
\setlength{\tabcolsep}{5pt}
\begin{tabular}{@{}llccc@{}}
\toprule
Protocol & Model & Accuracy & Macro F1 & AUC \\
\midrule
Single-stage four-class & TabPFN & 0.619 (0.062) & 0.627 (0.064) & 0.865 (0.027) \\
Single-stage four-class & Avg(R8) & 0.613 (0.032) & 0.618 (0.035) & 0.843 (0.014) \\
Single-stage four-class & Avg(13) & 0.613 (0.051) & 0.614 (0.047) & 0.838 (0.014) \\
Single-stage four-class & SVM & 0.594 (0.029) & 0.594 (0.029) & 0.843 (0.019) \\
Single-stage four-class & AG:NN & 0.600 (0.093) & 0.605 (0.094) & 0.837 (0.027) \\
\midrule
Two-stage cascade 2+2 & LR $\rightarrow$ TabPFN & 0.626 (0.018) & 0.631 (0.022) & 0.859 (0.025) \\
Two-stage cascade 2+2 & LR $\rightarrow$ SVM & 0.626 (0.049) & 0.634 (0.054) & 0.855 (0.022) \\
Two-stage cascade 2+2 & AG:LR $\rightarrow$ SVM & 0.632 (0.071) & 0.640 (0.073) & 0.853 (0.036) \\
Two-stage cascade 2+2 & LR $\rightarrow$ LR & 0.632 (0.043) & 0.641 (0.045) & 0.851 (0.018) \\
Two-stage cascade 2+2 & SVM $\rightarrow$ SVM & 0.626 (0.059) & 0.638 (0.061) & 0.851 (0.022) \\
\midrule
Two-stage cascade 1+3 & TabPFN $\rightarrow$ Avg(R8) & 0.619 (0.053) & 0.626 (0.060) & 0.851 (0.011) \\
Two-stage cascade 1+3 & AG:ETG $\rightarrow$ Avg(T7) & 0.626 (0.054) & 0.635 (0.056) & 0.843 (0.015) \\
Two-stage cascade 1+3 & AG:ETG $\rightarrow$ Avg(13) & 0.632 (0.067) & 0.637 (0.075) & 0.838 (0.015) \\
Two-stage cascade 1+3 & AG:ETG $\rightarrow$ Avg(R8) & 0.626 (0.063) & 0.631 (0.072) & 0.838 (0.014) \\
Two-stage cascade 1+3 & AG:RFE $\rightarrow$ Avg(13) & 0.626 (0.067) & 0.631 (0.075) & 0.832 (0.017) \\
\midrule
Three-stage cascade 2+2+2 & TabPFN $\rightarrow$ SVM $\rightarrow$ TabPFN & 0.658 (0.063) & 0.664 (0.066) & 0.869 (0.031) \\
Three-stage cascade 2+2+2 & AG:ETG $\rightarrow$ SVM $\rightarrow$ RF & 0.645 (0.079) & 0.651 (0.085) & 0.854 (0.026) \\
\textbf{Three-stage cascade 2+2+2} & \textbf{AG:ETG $\rightarrow$ SVM $\rightarrow$ TabPFN} & \textbf{0.665 (0.081)} & \textbf{0.670 (0.085)} & \textbf{0.851 (0.033)} \\
Three-stage cascade 2+2+2 & Avg(T7) $\rightarrow$ SVM $\rightarrow$ TabPFN & 0.652 (0.053) & 0.657 (0.056) & 0.845 (0.036) \\
Three-stage cascade 2+2+2 & AG:ETG $\rightarrow$ SVM $\rightarrow$ LR & 0.645 (0.091) & 0.653 (0.095) & 0.840 (0.038) \\
\bottomrule
\end{tabular}
\begin{tablenotes}[flushleft]\footnotesize
\item Values are mean (SD) over five folds of stratified cross-validation. Configurations are ordered by the sum of their ranks across accuracy, macro F1, and AUC (rank 1 = best per metric); each listed configuration ranks within the top five for at least two of the three metrics. AG, AutoGluon; NN, NeuralNetTorch; ETE, ExtraTreesEntr; ETG, ExtraTreesGini; RFE, RandomForestEntr; LR, logistic regression; AG:LR, AutoGluon LinearModel; RF, random forest; SVM, support vector machine (RBF); TabPFN, TabPFN transformer; Avg(13), simple-average ensemble of 13 base AutoGluon models; Avg(T7), ensemble of the seven tree models; Avg(R8), ensemble of one representative per model family.
\end{tablenotes}
\end{threeparttable}
% \end{revblock}
\end{table}

\section{Discussion}

In this cross-sectional eye-tracking study, depression symptom severity was associated with broad changes in how young participants sampled, organized, retrieved, and maintained visual information. The most consistent associations were lower gaze-allocation entropy, fewer and less diverse transitions between emotional faces, slower and less accurate orienting to a remembered location, and weaker gaze maintenance after target acquisition. These effects were large in standardized units and remained significant after correction across the prespecified primary feature set. In contrast, none of the emotion-specific attentional-bias features was associated with severity, and fixation-stability findings were limited. The pattern therefore points more strongly to distributed visual exploration and memory-guided control than to a generalized negative gaze bias.

The free-viewing findings extend prior reports of altered scanpath length, saccade amplitude, fixation behavior, and visual exploration in depression \cite{Takahashi2021,Wang2022,Gao2023,Zhang2022}. Lower first-pass coverage and entropy indicate that participants with greater symptoms sampled the available facial information less comprehensively and less evenly. Lower inter-AOI switch rate and transition diversity, together with greater tortuosity, further indicate altered organization of the scanpath. These measures should not be interpreted as direct readouts of a single cognitive mechanism. They may jointly reflect differences in task engagement, processing speed, spatial encoding, executive allocation, or psychomotor behavior, and the present design cannot distinguish among these explanations.

The absence of a significant F2 signal is clinically and theoretically informative. Meta-analyses of passive or sustained viewing have identified greater maintenance on negative material and reduced attention to positive material in depression \cite{Armstrong2012,Suslow2020}. Here, however, every emotional face was potentially relevant because the target category was revealed only after encoding. This task demand may have reduced spontaneous valence-driven allocation by encouraging participants to encode all locations. The null results do not exclude context-dependent affective bias. Rather, they show that, under a memory-relevant viewing requirement, symptom severity was expressed more consistently in coverage and transition organization than in negative-versus-positive allocation.

The recall findings suggest that the largest severity-related differences emerged when encoded information had to guide a goal-directed gaze response. Longer target-entry and target-fixation latencies, lower initial target-entry accuracy, and reduced dwell after target acquisition are consistent with less efficient memory-guided orienting and maintenance. Prior studies have linked depression to cue-detection deficits and impaired memory-guided allocation \cite{Chen2013,Nemeth2016}, and adolescent work has associated eye-movement characteristics with cognitive functioning \cite{Lu2024}. Nevertheless, the current task combines spatial memory, response selection, visual search, and oculomotor stabilization. The results should therefore be described as task-related behavioral correlates rather than evidence of isolated memory impairment or a specific neurocognitive pathway.

From a digital-medicine perspective, a brief, standardized eye-tracking task could complement clinical interviews by providing repeatable behavioral measurements that are difficult to obtain from symptom reports alone. The domain framework also provides a more interpretable alternative to treating a large collection of eye-movement variables as independent biomarkers. Potential applications could include longitudinal phenotyping, treatment-response research, or stratification in mechanistic studies. Such uses remain prospective. The study did not evaluate diagnostic discrimination, test--retest reliability, responsiveness to clinical change, patient acceptability, implementation burden, or added value beyond routine assessment. Specialized hardware, calibration requirements, stimulus licensing, and intersite standardization also limit immediate scalability.

% \begin{revblock}
The exploratory machine-learning results extend the statistical framework to a discriminative question. Among the four protocols, three-stage cascades outperformed two-stage and single-stage classification (best four-class accuracy 0.665, macro F1 0.670, AUC 0.869), suggesting that severity discrimination benefits from progressively finer hierarchical decisions rather than a single four-way decision. The exploratory two-class task (no/mild versus moderate/severe) was substantially easier (best accuracy 0.839, AUC 0.929), indicating that the coarse symptomatic boundary is well supported by these features, while the residual four-class errors concentrated near the moderate--severe boundary. This mirrors clinical practice, in which the moderate--severe distinction rests on threshold items of the HAMD-17 that are intrinsically less well defined than the distinction between no depression and symptomatic illness. Refining moderate-versus-severe discrimination, for example with richer severity ratings or longitudinal anchors, is therefore a key direction for future work.
% \end{revblock}

Several design features constrain the severity interpretation. The no-depression group was recruited from the community, whereas symptomatic participants were recruited from a clinical service, so the ordered gradient may partly encode differences in diagnostic status, referral pathway, medication exposure, illness chronicity, or testing context. Age and education also differed across groups. Although both were included as covariates, residual confounding remains possible, and adjusting simultaneously for highly correlated developmental variables may not fully recover comparability. A within-MDD continuous HAMD-17 analysis, propensity or overlap weighting, flexible age adjustment, and demographic-only benchmark models would strengthen the claim that the eye-movement features track symptom burden.

The HAMD-17 grouping also warrants caution. Thresholds create clinically recognizable strata but discard within-group information and impose equal spacing between categories. Moreover, HAMD-17 includes psychomotor items, raising the possibility that some eye-movement associations reflect overlap with clinician-rated motor behavior. Continuous HAMD-17 models should be emphasized, and sensitivity analyses should exclude psychomotor items or examine symptom dimensions. Anxiety was addressed only by covariate adjustment. Because depression and anxiety symptoms overlap, HAMA adjustment may either reduce confounding or remove clinically shared variance, and both interpretations should be considered.

Other limitations include the cross-sectional and single-center design, modest group sizes, exclusion of 48 participants on eye-tracking quality criteria, and incomplete reporting of medication, treatment, illness duration, comorbidity, sleep, and visual characteristics. The sample age range of 14--20 years extends beyond conventional definitions of adolescence and should be described as adolescents and young adults unless a prespecified alternative definition is justified. The fixed trial sequence and progressively shorter display duration may introduce order, fatigue, learning, or target-emotion confounding. The requirement for valid data in all 12 trials may also select participants with better task adherence. 
The machine-learning analysis was exploratory, used a single five-fold split of a small sample, and reported best-of-many configurations (441 two-stage or 9261 three-stage), so its best values are optimistic upper bounds on generalizable performance.
External validation, trial-level mixed-effects analyses, calibration and reliability studies, and prospective longitudinal evaluation are required.

In conclusion, the combined encoding and recall paradigm identified a coherent severity-related pattern across visual sampling, scanpath organization, memory-guided acquisition, and target maintenance, while providing little evidence for generalized negative gaze bias; an exploratory hierarchical classifier reached the best four-class separation (accuracy 0.665, macro F1 0.670, AUC 0.869). 
These findings support eye tracking as a candidate research tool for task-based behavioral phenotyping of depression symptom severity. They do not establish a depression-specific mechanism, diagnostic biomarker, or clinically deployable assessment.

\section{Methods}

\subsection{Study design and reporting}

This was a single-center, cross-sectional observational eye-tracking study. Patients with major depressive disorder were recruited from the Department of Psychology, Shandong Provincial Hospital Affiliated to Shandong First Medical University, and participants without depression were recruited from local communities. 

\subsection{Participants}

Eligible participants were aged 14--20 years, had completed at least primary education, and had normal or corrected-to-normal vision. Patients were required to meet DSM-5 criteria for major depressive disorder in a clinical interview conducted by an experienced psychiatrist and to have a HAMD-17 score above the normal range. Community participants were required to have a HAMD-17 score within the normal range and no personal or family history of depression.

Exclusion criteria were a history of neurological disease, traumatic brain injury, epilepsy, or another organic brain disorder; a current or previous major psychiatric disorder other than major depressive disorder, including schizophrenia, bipolar disorder, or substance-use disorder; intellectual disability or cognitive impairment sufficient to interfere with task completion; color blindness or another ocular condition affecting eye tracking; and missing key clinical or behavioral data or inadequate eye-tracking quality. 

\subsection{Clinical assessment and symptom-severity strata}

Depressive symptoms were rated using the HAMD-17, and anxiety symptoms were rated using the HAMA \cite{Shear2001}. Both scales were administered before eye tracking by trained clinical assessors. The HAMD-17 was used as the principal measure of depression symptom severity \cite{Khan2010,Shafer2006,Seemuller2023}. 

Participants were grouped using HAMD-17 thresholds of 0--7, 8--16, 17--23, and 24 or higher, corresponding to no, mild, moderate, and severe symptom strata \cite{Zimmerman2013,Zimmerman2016,Lecrubier2007}. These labels describe HAMD-17 symptom strata and should not be interpreted as independent diagnostic categories. 

\subsection{Ethics and consent}

The study complied with the Declaration of Helsinki and was approved by the Ethics Committee for Biomedical Research Involving Human Subjects, Shandong Provincial Hospital Affiliated to Shandong First Medical University (R202411120385). Written informed consent was obtained from adult participants. For participants younger than 18 years, written permission was obtained from a parent or legal guardian. 

\subsection{Eye-tracking apparatus}

Eye movements were recorded using a Tobii Pro Fusion system (Tobii Technology, Stockholm, Sweden) at 250 Hz and exported using Tobii Pro Lab version 25.23. Participants completed a nine-point calibration before the task. Fixations were identified using the Tobii Pro Lab fixation filter with a minimum duration of 100 ms. 

\subsection{Stimuli}

The task used 36 facial images from the Chinese Facial Affective Picture System, with equal numbers of male and female identities and 12 images in each of the positive, neutral, and negative categories \cite{Gong2011}. Normative mean (SD) valence and arousal ratings were 2.75 (0.46) and 5.97 (0.78) for negative faces, 4.42 (0.32) and 3.97 (0.69) for neutral faces, and 6.56 (0.59) and 5.28 (0.71) for positive faces. Images were converted to grayscale and standardized for size and resolution. Each image appeared three times, with pseudo-randomized screen position across displays. 

\subsection{Emotional-spatial free-viewing and memory-guided recall paradigm}

\begin{figure}[!t]
\centering
\includegraphics[width=0.93\textwidth]{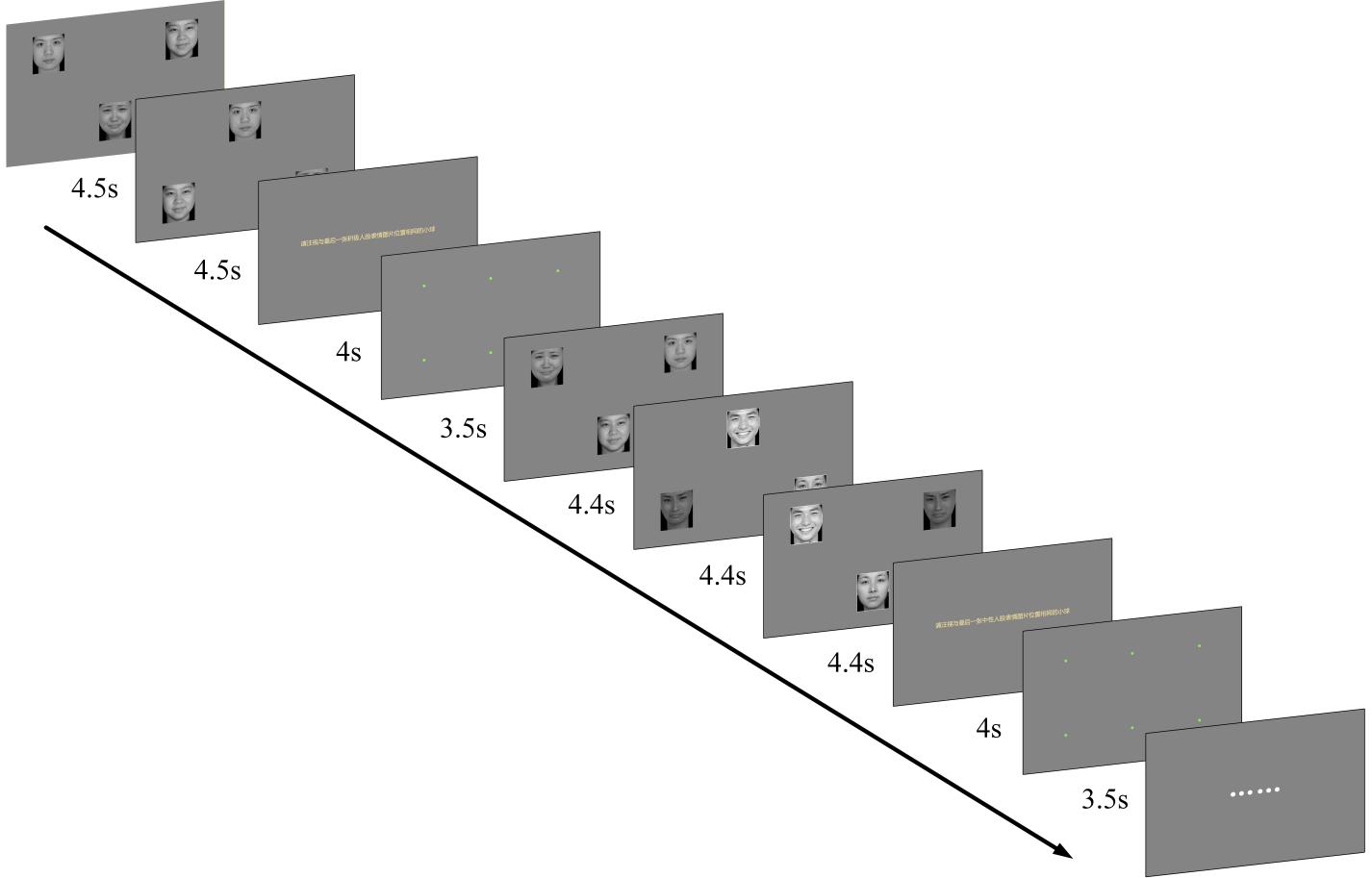}
\caption{\textbf{Emotional-spatial free-viewing and memory-guided recall paradigm.} Each trial comprised two to four displays containing positive, neutral, and negative faces, followed by an emotional-category cue and a six-location recall probe. The cue referred to the final face display immediately preceding it. }
\label{fig:paradigm}
\end{figure}

The paradigm paired emotional-spatial free-viewing with a subsequent memory-guided recall response (Fig.~\ref{fig:paradigm}). During free-viewing, each display contained one positive, one neutral, and one negative face at three locations in an upright or inverted triangular arrangement. Participants knew that a subsequent cue could request any emotional category, but the target was not revealed during viewing. The task therefore required simultaneous sampling of emotional faces and encoding of their spatial locations rather than unconstrained passive viewing.

After the free-viewing displays in a trial, a cue indicated the emotional category to be recalled from the final display. A probe screen then showed six fixation points corresponding to all possible face locations. Participants were instructed to direct gaze to the remembered target location and maintain fixation until the next trial. The paradigm included 12 trials, with each emotion cued four times. Trials contained two, three, or four free-viewing displays in a repeated cycle, yielding 36 displays. Display duration decreased from 4.5 s to 3.4 s in 0.1-s steps across trials. The cue was presented for 4.0 s and the probe for 3.5 s. The target-emotion sequence was pseudo-randomized but identical across participants. Stimuli were displayed on a 24-inch, 1920 $\times$ 1080 monitor at a viewing distance of approximately 60--80 cm.

The identical sequence and progressively changing display duration standardized exposure but also coupled trial order with task difficulty.

\subsection{Gaze preprocessing and quality control}

Raw gaze samples were restricted to valid samples and segmented according to task phase and trial. Gaze coordinates were mapped to prespecified AOIs. Free-viewing features were aggregated from display to trial and then participant level. Recall features were computed by trial relative to the target AOI and then aggregated to participant level. Participants were included only if their overall valid-sample ratio was at least 80\% and all 12 trials were usable.

A recall trial was classified as correct when gaze entered and remained within the target AOI according to a prespecified offline rule.

\subsection{Task-informed eye-movement feature framework}

Features were prespecified according to the task phase and the aspect of gaze behavior they were intended to characterize. Free-viewing features were grouped into visual scanning coverage (F1), emotion-related attentional bias (F2), and gaze transition dynamics (F3). Recall features were grouped into memory-guided target acquisition (R1), target-area gaze maintenance (R2), and fixation stability (R3) (Table~\ref{tab:domains}). These domains are task-informed interpretations and not direct measurements of isolated cognitive mechanisms. Operational definitions and calculation procedures for all features are provided in Table~\ref{tab:s1_feature_definitions}.

\begin{table}[!t]
\centering
\begin{threeparttable}
\caption{\textbf{Prespecified eye-movement feature domains.}}
\label{tab:domains}
\small
\begin{tabularx}{\textwidth}{p{0.12\textwidth}p{0.22\textwidth}Y Y}
\toprule
Task & Domain & Task-related interpretation & Primary features \\
\midrule
Free-viewing & F1 Visual scanning coverage & Completeness and balance of visual sampling across the three emotional face AOIs & Time to full AOI coverage; first-pass AOI coverage ratio; AOI allocation entropy \\
Free-viewing & F2 Emotion-related attentional bias & Valence-related allocation and maintenance, particularly negative maintenance and delayed disengagement & Negative--positive gaze allocation bias; negative-AOI visit duration; negative stickiness after first hit \\
Free-viewing & F3 Gaze transition dynamics & Spatial and temporal organization of gaze across and within emotional AOIs & Inter-AOI switch rate; AOI transition diversity; transition tortuosity; within-AOI gaze dispersion \\
Recall & R1 Memory-guided target acquisition & Efficiency and accuracy of orienting to the cued location using encoded spatial information & First target-entry latency; initial target-entry accuracy; first target-fixation latency \\
Recall & R2 Target-area gaze maintenance & Preferential and sustained gaze on the recalled target after acquisition & Target dwell proportion; target lock-on ratio; post-hit off-target ratio \\
Recall & R3 Fixation stability & Local spatial stability of gaze within the target area & Target-area fixation dispersion; within-target fixation step size; target-area stability ellipse area \\
\bottomrule
\end{tabularx}
\begin{tablenotes}[flushleft]\footnotesize
\item AOI, area of interest. 
\end{tablenotes}
\end{threeparttable}
\end{table}

\subsection{Statistical analysis}

Analyses were conducted at participant level. Continuous demographic and clinical variables were compared across the four HAMD-17 groups using Kruskal--Wallis tests, and sex was compared using a chi-square test. Each primary eye-movement feature was standardized before model fitting. For feature $Y_j$, the primary model was
\begin{equation}
Y_{ij}^{(z)} = \beta_0 + \beta_1 G_i + \beta_2 \mathrm{Age}_i + \beta_3 \mathrm{Sex}_i + \beta_4 \mathrm{Education}_i + \epsilon_i,
\end{equation}
where $G_i$ was coded 0, 1, 2, or 3 for the ordered HAMD-17 strata. HC3 heteroscedasticity-consistent standard errors were used. Thus, $\beta_1$ represents the SD change in the feature associated with a one-level increase in symptom-severity stratum.

The primary family contained 19 prespecified features across six domains. Benjamini--Hochberg false-discovery-rate correction was applied across these 19 tests, and $\qfdr<0.05$ was considered statistically significant. All tests were two-sided. The 25 supplementary features were analyzed as supportive evidence. The  Benjamini--Hochberg false-discovery-rate correction was applied across the 25 supplementary features as a separate correction family.

Dimensional analyses examined Spearman correlations and linear regressions using continuous HAMD-17 scores, adjusted for age, sex, and education. Sensitivity analyses additionally adjusted for HAMA score and repeated the primary models after values with feature-specific $|z|>3$ were set to missing. 

% \begin{revblock}
\subsection{Machine-learning classification}
The machine-learning analysis was exploratory and used the same 155 participants and HAMD-17 severity strata as the statistical analysis. Inputs were the 44 candidate eye-movement features (25 free-viewing and 19 recall features, including the 19 prespecified primary features) plus age, sex, and education (47 inputs). The severity label, raw HAMD-17 and HAMA scores, and participant identifiers were excluded from all model inputs. The single missing value of one recall feature in three participants was imputed with the training-fold mean.

We evaluated 21 classifiers: logistic regression (L2 penalty, maximum 5000 iterations), random forest (500 trees), radial-basis-function support vector machine with probability calibration, the TabPFN transformer in its default configuration, fourteen AutoGluon 1.1.1 configurations (LightGBM, LightGBM extra-trees, large LightGBM, XGBoost, CatBoost, two random-forest, two extra-trees, two linear, two k-nearest-neighbour, and one neural-network model, plus a weighted ensemble fitted with a 20\% holdout of each training fold, refit on the full fold, without stacking or bagging), and three simple-average ensembles of the out-of-fold class probabilities of the 13 base AutoGluon models, of the seven tree models, or of one representative per model family (eight models).

Four protocols were tested. Single-stage: direct four-class classification of the severity strata. Two-stage 2+2: a first binary classifier separated no/mild from moderate/severe strata, and two second-stage binary classifiers separately split no from mild and moderate from severe. Two-stage 1+3: a first binary classifier separated no depression from the symptomatic strata, followed by a three-class classifier for mild, moderate, and severe. Three-stage 2+2+2: the 2+2 first stage followed by both second-stage binary classifiers. In cascade protocols, second- and third-stage models were trained, within each fold, on the true-branch training subsets, and test-fold participants were routed by the previous stage's predicted probabilities; the four stratum probabilities were the products of the stage probabilities along each path of the cascade tree (for the three-stage protocol, $(1-p_1)(1-p_2)$, $(1-p_1)p_2$, $p_1(1-p_2)(1-p_3)$, and $p_1(1-p_2)p_3$, where $p_1$, $p_2$, and $p_3$ are the first-stage moderate/severe, second-stage mild, and third-stage severe probabilities), and the predicted stratum was the argmax. All protocols used the same five-fold cross-validation (stratified on the four severity labels, shuffled, fixed seed), with 124 participants for training and 31 for testing in each fold; the exploratory two-class task was stratified on its two labels. Imputation and, for the three baseline classifiers, standardization, were refit within each training fold. Accuracy, macro-averaged F1, and one-vs-rest macro-averaged AUC were reported as the mean (sample SD) over folds. Analyses used Python 3.8.20, scikit-learn 1.3.2, AutoGluon 1.1.1, and TabPFN 8.4.0 on an NVIDIA RTX 5880 Ada Generation GPU.
% \end{revblock}

\subsection{Data availability}

Data supporting the findings of this study are not publicly available because they contain sensitive behavioral and clinical information from adolescent participants. Access to these data is restricted by ethical and institutional requirements.

\subsection{Code availability}

The custom code used for eye tracking data preprocessing, model training, and evaluation is not publicly available but may be made available to qualified researchers upon reasonable request to the corresponding authors, subject to institutional approval.

\section{Acknowledgements}

This work was supported in part by the National Key Research and Development Program of China under Grant 2023YFB4706100, the National Natural Science Foundation of China under Grant 62573271, the Major Basic Research Project of Shandong Provincial Natural Science Foundation under Grant ZR2025ZD24, and the Chinese University of Hong Kong Vice-Chancellor Early Career Professorship Scheme under Grant 441.

\section{Author contributions}

Youxiang Sun: Writing -- original draft, Writing -- review and editing, Conceptualization, Data curation. Yujie Nie: Writing -- original draft, Writing -- review and editing. Zhuoru Wu: Writing -- original draft, Writing -- review and editing, Formal analysis, Visualization. Mengyuan Yu: Writing -- original draft, Writing -- review and editing, Investigation, Data curation. Wenkong Wang: Writing -- original draft, Writing -- review and editing. Xiang Li: Writing -- original draft, Writing -- review and editing. Yumei Wang: Writing -- original draft, Writing -- review and editing, Supervision, Data curation. Xin Ma: Writing -- original draft, Writing -- review and editing, Supervision, Funding acquisition. Lizhou Fan: Writing -- original draft, Writing -- review and editing, Conceptualization, Supervision, Funding acquisition.

\section{Competing interests}
The authors declare no competing interests.

\clearpage

\section{Supplementary information}

\setcounter{figure}{0}
\renewcommand{\thefigure}{S\arabic{figure}}

\begin{figure}[htbp]
\centering
\includegraphics[width=\textwidth]{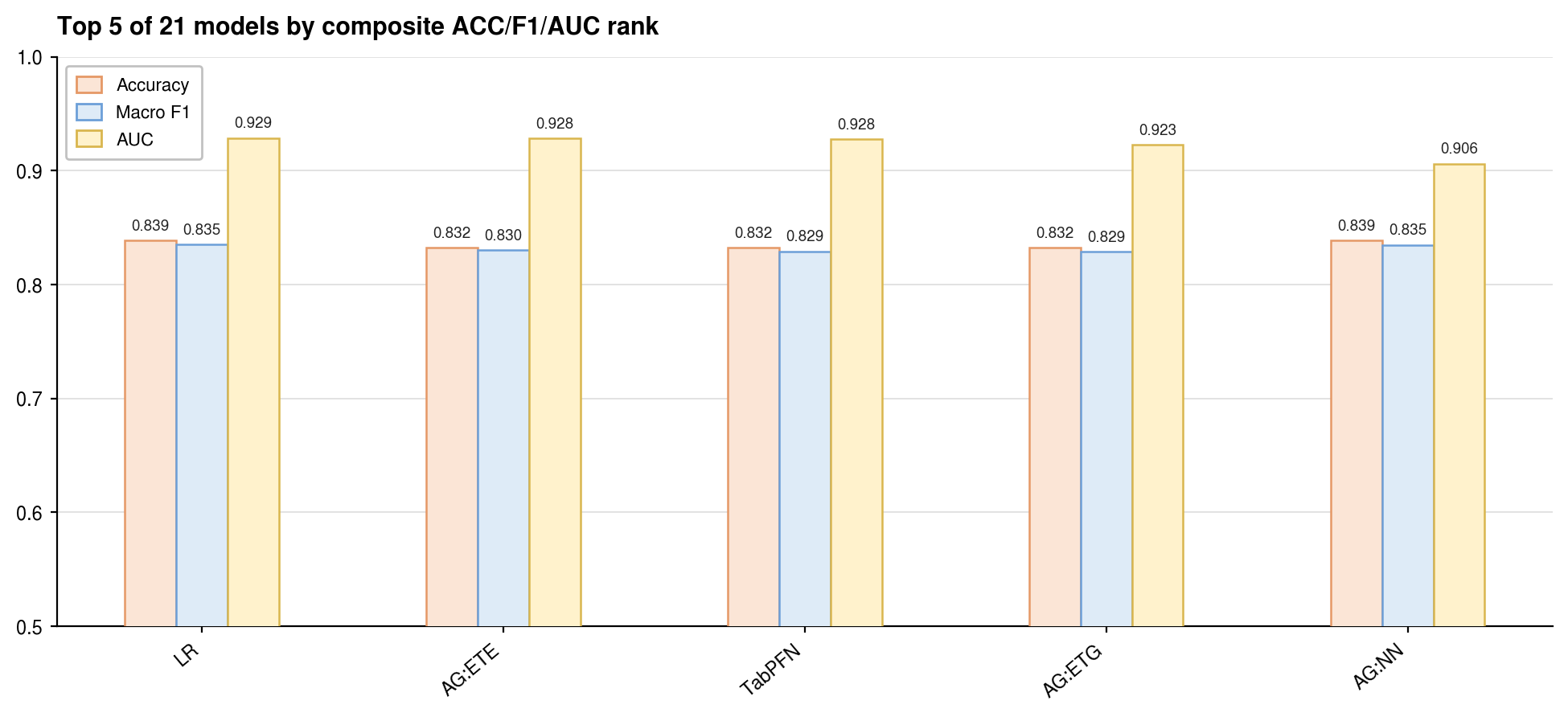}
\caption{\textbf{Top-5 classifiers for the exploratory two-class task (no/mild versus moderate/severe strata).} Models are ordered by the sum of their ranks across accuracy, macro F1, and AUC (top 5 shown, left to right); grouped bars show the three metric values of each model. Values are means over five folds of stratified cross-validation. Abbreviations as in Fig.~\ref{fig:ml_settings}.}
\label{fig:ml_binary}
\end{figure}
\clearpage

% ==========================================================
% S1 as ONE continuous full-width table over multiple pages.
% The surrounding document remains two-column. LaTeX's
% longtable needs full-width (one-column) pages while it runs.
% ==========================================================
% Number this as Supplementary Table S1.
\setcounter{table}{0}
\renewcommand{\thetable}{S\arabic{table}}
\begingroup
\sloppy
\emergencystretch=2em
\fontsize{7.5pt}{8.5pt}\selectfont
\setlength{\tabcolsep}{1.5pt}
\renewcommand{\arraystretch}{0.92}
\setlength{\extrarowheight}{0pt}
\setlength{\LTpre}{3pt}
\setlength{\LTpost}{3pt}
\begin{longtable}{@{}
    P{0.06\textwidth}
    P{0.24\textwidth}
    P{0.39\textwidth}
    P{0.28\textwidth}
@{}}

\caption{\textbf{Operational definitions and interpretations of extracted eye-movement features.}}
\label{tab:s1_feature_definitions}\\
\toprule
Domain & Feature & Operational calculation & Interpretation \\
\midrule
\endfirsthead
\multicolumn{4}{@{}l}{\textbf{Supplementary Table S1 (continued)}}\\[3pt]
\toprule
Domain & Feature & Operational calculation & Interpretation \\
\midrule
\endhead
\midrule
\multicolumn{4}{r@{}}{\footnotesize Continued on next page}\\
\endfoot
\bottomrule
\endlastfoot
F1 & Complete AOI coverage proportion & For each free-viewing display, code whether gaze entered all three emotional face AOIs at least once; average this binary value across displays. & Completeness of emotional AOI sampling during encoding; higher values indicate more complete coverage. \\
F1 & AOI allocation entropy & Compute Shannon entropy of dwell-time proportions across negative, neutral, and positive AOIs for each display; average across displays. & Balance of gaze allocation across emotional AOIs; higher values indicate more even sampling. \\
F1 & AOI allocation imbalance & Compute the L1 distance between the observed dwell-time distribution across the three emotional AOIs and a uniform distribution of 1/3, 1/3, 1/3; average across displays. & Deviation from balanced AOI allocation; higher values indicate more uneven sampling. \\
F1 & First-pass AOI coverage & During the first pass through AOIs before the first revisit, count the number of unique emotional AOIs visited and divide by three; average across displays. & Breadth of initial visual sampling; higher values indicate broader first-pass coverage. \\
F1 & Time to full AOI coverage & Compute the latency from display onset until all three emotional AOIs have been entered at least once; average across displays with full coverage. & Efficiency of completing emotional AOI coverage; higher values indicate slower coverage. \\
F2 & Negative-AOI disengagement latency & After the first entry into the negative AOI, compute the duration until gaze first leaves the negative AOI; average across displays with a negative-AOI hit. & Disengagement from negative information; higher values indicate slower disengagement from the negative AOI. \\
F2 & High-low load change in negative-positive gaze allocation bias & Compute negative-positive gaze allocation bias separately for high-load displays with four stimuli and low-load displays with two stimuli, then subtract low-load from high-load values. & Load-related change in negative versus positive gaze allocation; positive values indicate stronger negative bias under higher encoding load. \\
F2 & Negative AOI visit duration & Identify continuous visits to the negative AOI and compute the mean duration of these visits; average across displays. & Maintenance of gaze on negative emotional information; higher values indicate longer negative-AOI visits. \\
F2 & Negative-neutral gaze allocation bias & Compute the difference between the negative AOI dwell proportion and the neutral AOI dwell proportion for each display; average across displays. & Relative allocation of gaze to negative versus neutral stimuli; positive values indicate greater negative than neutral allocation. \\
F2 & Negative-positive gaze allocation bias & Compute the difference between the negative AOI dwell proportion and the positive AOI dwell proportion for each display; average across displays. & Relative allocation of gaze to negative versus positive stimuli; positive values indicate greater negative than positive allocation. \\
F2 & Negative stickiness after first hit & After the first hit on the negative AOI, compute the proportion of the remaining viewing time spent on the negative AOI; average across displays with a negative-AOI hit. & Post-engagement maintenance on negative information; higher values indicate stronger negative stickiness after first hit. \\
F2 & Late-early change in negative-positive gaze allocation bias & Compute negative-positive gaze allocation bias separately for late and early task phases, then subtract early-phase from late-phase values. & Task-phase change in negative versus positive gaze allocation; positive values indicate increasing negative bias over time. \\
F2 & Late-early change in negative stickiness & Compute negative stickiness after first hit separately for late and early task phases, then subtract early-phase from late-phase values. & Task-phase change in negative attention stickiness; positive values indicate increasing negative stickiness over time. \\
F2 & Negative AOI revisit count & Count repeated entries into the negative AOI after the initial negative-AOI visit; average across displays. & Recurrent return to negative information; higher values indicate more negative-AOI revisits. \\
F3 & Inter-AOI switch rate & Count transitions between different emotional AOIs and divide by valid free-viewing time in seconds; average across displays. & Frequency of switching among emotional AOIs; higher values indicate more frequent inter-AOI exploration. \\
F3 & AOI transition entropy & Compute Shannon entropy over observed AOI-to-AOI transition categories for each display; average across displays. & Diversity of inter-AOI transition patterns; higher values indicate more varied transition sequences. \\
F3 & Normalized gaze convex-hull area & Compute the convex-hull area covering valid gaze samples during free viewing and normalize it by the fixed stimulus/display coordinate scale; average across displays. & Global spatial spread of visual exploration; higher values indicate a wider exploration field. \\
F3 & Mean saccade amplitude & Compute the mean amplitude of saccadic gaze shifts during free viewing and normalize by the fixed stimulus/display coordinate scale; average across displays. & Spatial extent of saccadic exploration; higher values indicate larger gaze shifts. \\
F3 & Normalized scanpath length & Sum consecutive gaze-step distances during free viewing and normalize by the fixed stimulus/display coordinate scale; average across displays. & Overall amount of gaze movement during encoding; higher values indicate longer scanpaths. \\
F3 & AOI transition duration & For inter-AOI transitions, compute the elapsed time from leaving one AOI to entering the next AOI; average across transitions and displays. & Temporal cost of moving between AOIs; higher values indicate slower inter-AOI transitions. \\
F3 & Outside-AOI transition pause duration & During inter-AOI transitions, compute time spent outside all emotional AOIs before the next AOI is entered; average across transitions and displays. & Off-AOI pause during transitions; higher values indicate longer non-AOI dwell while shifting between emotional faces. \\
F3 & Transition tortuosity & For inter-AOI transitions, divide the actual gaze path length by the straight-line distance between the source and destination AOI centers; average across transitions. & Geometric efficiency of inter-AOI transitions; higher values indicate more curved or indirect transition trajectories. \\
F3 & Within-AOI gaze dispersion & Within AOI visits, compute the root-mean-square distance of gaze samples from the local AOI-visit centroid and normalize by AOI/display scale; average across visits and displays. & Local spatial dispersion of gaze within AOIs; higher values indicate less concentrated within-AOI viewing. \\
F3 & Within-AOI gaze step size & Within AOI visits, compute the mean distance between consecutive gaze samples and normalize by AOI/display scale; average across visits and displays. & Local within-AOI gaze wandering; higher values indicate greater moment-to-moment movement within AOIs. \\
R1 & Target-hit proportion & Code whether gaze ever entered the cued target AOI on each recall trial and average across trials. & Whether the recalled target was successfully reached; higher values indicate more frequent target hits. \\
R1 & Initial target-entry accuracy & For each recall trial, identify the first candidate AOI entered and code whether it was the cued target; average across trials. & Accuracy of initial memory-guided orienting; higher values indicate more accurate first target entry. \\
R1 & Pre-target non-target AOI count & Before first target entry, count entries into non-target candidate AOIs; average across recall trials. & Pre-target search errors or distractor sampling; higher values indicate more non-target exploration before target acquisition. \\
R1 & Pre-target search path length & Before first target entry, sum consecutive gaze-step distances and normalize by the fixed stimulus/display coordinate scale; average across trials. & Spatial search effort before acquiring the target; higher values indicate longer pre-target search paths. \\
R1 & Strict target-fixation latency & Compute the latency from recall onset to the first strict fixation event within the target AOI; average across trials with strict target fixation. & Time required to form stable fixation on the recalled target; higher values indicate delayed stable target fixation. \\
R1 & Target-entry latency & Compute the latency from recall onset to the first entry into the cued target AOI; average across trials with target entry. & Speed of memory-guided target acquisition; higher values indicate delayed target entry. \\
R2 & Longest continuous target run & For each recall trial, compute the longest continuous dwell episode within the target AOI; average across trials. & Sustained continuous gaze maintenance on the recalled target; higher values indicate longer uninterrupted target holding. \\
R2 & Off-target dwell proportion & Compute the proportion of recall viewing time spent in non-target candidate AOIs; average across trials. & Allocation to competing non-target locations; higher values indicate stronger off-target allocation. \\
R2 & Post-hit off-target ratio & After first target entry, compute the proportion of the remaining recall time spent in non-target AOIs; average across trials with target entry. & Off-target drift after target acquisition; higher values indicate weaker maintenance of the recalled target. \\
R2 & Target dwell proportion & Compute the proportion of recall viewing time spent within the cued target AOI; average across trials. & Preferential allocation to the recalled target; higher values indicate stronger target prioritization. \\
R2 & Target lock-on ratio & After first target entry, compute the proportion of the remaining recall time spent within the target AOI; average across trials with target entry. & Maintenance of gaze on the target after acquisition; higher values indicate stronger target lock-on. \\
R2 & Target re-entry count after first hit & After first target entry, count subsequent re-entries into the target AOI following exits; average across trials. & Return behavior after leaving the target; higher values indicate more repeated target reacquisition. \\
R3 & Candidate target-area fixation dispersion & Within any candidate AOI during recall, compute the root-mean-square distance of strict fixation samples from their local centroid and normalize by AOI/display scale; average across candidate AOIs/trials. & General fixation dispersion within candidate target areas, including target and non-target AOIs; higher values indicate lower local stability. \\
R3 & Candidate target-area fixation step size & Within any candidate AOI during recall, compute the mean consecutive step size of strict fixation samples and normalize by AOI/display scale; average across candidate AOIs/trials. & General fixation wandering within candidate target areas; higher values indicate less stable local fixation. \\
R3 & Target-area stability ellipse area & For strict fixation samples within the target AOI, compute a fixation-stability ellipse area from the horizontal and vertical gaze dispersion and normalize by AOI/display scale; average across trials. & Spatial spread of target-area fixation; higher values indicate lower fixation stability. \\
R3 & Strict target-fixation count & Count strict fixation events within the target AOI on each recall trial; average across trials. & Frequency of stable target fixations; higher values indicate more target fixation episodes. \\
R3 & Total strict target-fixation duration & Sum the duration of strict fixation events within the target AOI on each recall trial; average across trials. & Total stable fixation time on the recalled target; higher values indicate more sustained stable target viewing. \\
R3 & Target-area fixation dispersion & For strict fixation samples within the target AOI, compute the root-mean-square distance from the local fixation centroid and normalize by AOI/display scale; average across trials. & Local fixation dispersion within the target area; higher values indicate lower fixation stability. \\
R3 & Within-target fixation step size & For strict fixation samples within the target AOI, compute the mean distance between consecutive samples and normalize by AOI/display scale; average across trials. & Moment-to-moment instability within target fixation; higher values indicate greater local fixation wandering. \\
R3 & Total strict target-fixation duration & Sum the duration of strict fixation events within the target AOI on each recall trial; average across trials. & Total stable fixation time on the recalled target; higher values indicate more sustained stable target viewing. \\
\end{longtable}
\endgroup
\clearpage

\begingroup

% Compact formatting
\fontsize{7pt}{8pt}\selectfont
\setlength{\tabcolsep}{1.5pt}
\renewcommand{\arraystretch}{0.92}
\setlength{\extrarowheight}{0pt}

% Reduce whitespace around longtable
\setlength{\LTpre}{0pt}
\setlength{\LTpost}{0pt}

\begin{longtable}{@{}
    P{0.065\textwidth}
    P{0.48\textwidth}
    C{0.255\textwidth}
    C{0.08\textwidth}
    C{0.08\textwidth}
@{}}

\caption{\textbf{Supplementary results from ordered-severity models.}}
\label{tab:s2_ordered_severity}\\

\toprule
Domain
& Supplementary feature
& Standardized $\beta$ (95\% CI)
& $p$
& $q_{\mathrm{FDR}}$ \\
\midrule
\endfirsthead

\multicolumn{5}{@{}l}{%
\textbf{Supplementary Table \thetable\ (continued)}
}\\

\toprule
Domain
& Supplementary feature
& Standardized $\beta$ (95\% CI)
& $p$
& $q_{\mathrm{FDR}}$ \\
\midrule
\endhead

\midrule
\multicolumn{5}{r@{}}{%
\fontsize{7pt}{8pt}\selectfont Continued on next page
}\\
\endfoot

\bottomrule
\endlastfoot

% ===================== F1 =====================

F1 & Complete AOI coverage proportion
& $-0.538\,(-0.679, -0.398)$ & $<0.001$ & $<0.001$ \\

F1 & AOI allocation imbalance
& $0.603\,(0.477, 0.729)$ & $<0.001$ & $<0.001$ \\

% ===================== F2 =====================

F2 & Negative-neutral gaze allocation bias
& $0.118\,(-0.031, 0.267)$ & $0.120$ & $0.157$ \\

F2 & Negative AOI revisit count
& $-0.215\,(-0.368, -0.062)$ & $0.006$ & $0.009$ \\

F2 & Negative-AOI disengagement latency
& $-0.072\,(-0.238, 0.093)$ & $0.392$ & $0.466$ \\

F2 & Late-early change in negative-positive gaze allocation bias
& $-0.129\,(-0.295, 0.036)$ & $0.126$ & $0.157$ \\

F2 & Late-early change in negative stickiness
& $-0.006\,(-0.157, 0.144)$ & $0.933$ & $0.933$ \\

F2 & High-low load change in negative-positive gaze allocation bias
& $0.019\,(-0.143, 0.181)$ & $0.822$ & $0.857$ \\

F2 & High-low load change in negative stickiness
& $-0.061\,(-0.218, 0.097)$ & $0.450$ & $0.512$ \\

% ===================== F3 =====================

F3 & Normalized scanpath length
& $0.326\,(0.182, 0.471)$ & $<0.001$ & $<0.001$ \\

F3 & Normalized gaze convex-hull area
& $0.146\,(-0.015, 0.306)$ & $0.075$ & $0.104$ \\

F3 & AOI transition duration
& $0.409\,(0.255, 0.563)$ & $<0.001$ & $<0.001$ \\

F3 & Outside-AOI transition pause duration
& $0.288\,(0.159, 0.416)$ & $<0.001$ & $<0.001$ \\

F3 & Within-AOI gaze step size
& $0.308\,(0.168, 0.449)$ & $<0.001$ & $<0.001$ \\

F3 & Mean saccade amplitude
& $-0.348\,(-0.493, -0.204)$ & $<0.001$ & $<0.001$ \\

% ===================== R1 =====================

R1 & Target-hit proportion
& $-0.577\,(-0.690, -0.463)$ & $<0.001$ & $<0.001$ \\

R1 & Pre-target non-target AOI count
& $0.442\,(0.318, 0.566)$ & $<0.001$ & $<0.001$ \\

R1 & Pre-target search path length
& $0.550\,(0.416, 0.684)$ & $<0.001$ & $<0.001$ \\

% ===================== R2 =====================

R2 & Off-target dwell proportion
& $0.390\,(0.269, 0.511)$ & $<0.001$ & $<0.001$ \\

R2 & Longest continuous target run
& $-0.516\,(-0.646, -0.386)$ & $<0.001$ & $<0.001$ \\

R2 & Target re-entry count after first hit
& $-0.050\,(-0.190, 0.090)$ & $0.483$ & $0.525$ \\

% ===================== R3 =====================

R3 & Strict target-fixation count
& $-0.215\,(-0.360, -0.070)$ & $0.004$ & $0.006$ \\

R3 & Total strict target-fixation duration
& $-0.683\,(-0.785, -0.580)$ & $<0.001$ & $<0.001$ \\

R3 & Candidate target-area fixation dispersion
& $0.251\,(0.100, 0.402)$ & $0.001$ & $0.002$ \\

R3 & Candidate target-area fixation step size
& $0.143\,(-0.008, 0.294)$ & $0.063$ & $0.093$ \\

\end{longtable}

\vspace{-5pt}
\noindent
\begin{minipage}{\textwidth}
\fontsize{7pt}{8pt}\selectfont
Standardized coefficients were obtained from ordered-severity regression
models adjusted for age, gender, and years of education, using
heteroscedasticity-consistent robust standard errors.
FDR correction was applied across all 25 features.
Supplementary features were used to evaluate convergent patterns
within the same mechanism-informed domain.
FDR, false discovery rate; CI, confidence interval.
\end{minipage}

\endgroup
\clearpage

\begingroup
\fontsize{7pt}{8pt}\selectfont
\setlength{\tabcolsep}{1.5pt}
\renewcommand{\arraystretch}{0.92}
\setlength{\extrarowheight}{0pt}

% Reduce whitespace
\setlength{\LTpre}{0pt}
\setlength{\LTpost}{0pt}

\begin{longtable}{@{}
    P{0.055\textwidth}
    P{0.255\textwidth}
    C{0.075\textwidth}
    C{0.070\textwidth}
    C{0.075\textwidth}
    C{0.235\textwidth}
    C{0.070\textwidth}
    C{0.075\textwidth}
@{}}

\caption{\textbf{Associations between continuous HAMD-17 scores and supplementary eye-movement features.}}
\label{tab:s3_continuous_hamd17}\\

\toprule
Feature domain
& Supplementary feature
& \multicolumn{3}{c}{Spearman correlation}
& \multicolumn{3}{c}{Adjusted regression model} \\

\cmidrule(lr){3-5}
\cmidrule(lr){6-8}

& & $\rho_s$ & $p$ & $q_{\mathrm{FDR}}$
& $\beta$ (95\% CI) & $p$ & $q_{\mathrm{FDR}}$ \\

\midrule
\endfirsthead

\multicolumn{8}{@{}l}{
\textbf{Supplementary Table \thetable\ (continued)}
}\\

\toprule
Feature domain
& Supplementary feature
& \multicolumn{3}{c}{Spearman correlation}
& \multicolumn{3}{c}{Adjusted regression model} \\

\cmidrule(lr){3-5}
\cmidrule(lr){6-8}

& & $\rho_s$ & $p$ & $q_{\mathrm{FDR}}$
& $\beta$ (95\% CI) & $p$ & $q_{\mathrm{FDR}}$ \\

\midrule
\endhead

\midrule
\multicolumn{8}{r@{}}{
\fontsize{7pt}{8pt}\selectfont
Continued on next page
}\\
\endfoot

\bottomrule
\endlastfoot

% ===================== F1 =====================

F1 & Complete AOI coverage proportion
& $-0.600$ & $<0.001$ & $<0.001$
& $-0.490\,(-0.627, -0.354)$
& $<0.001$ & $<0.001$ \\

F1 & AOI allocation imbalance
& $0.614$ & $<0.001$ & $<0.001$
& $0.550\,(0.418, 0.682)$
& $<0.001$ & $<0.001$ \\

% ===================== F2 =====================

F2 & Negative--neutral gaze allocation bias
& $0.151$ & $0.060$ & $0.088$
& $0.113\,(-0.034, 0.261)$
& $0.132$ & $0.171$ \\

F2 & Negative AOI revisit count
& $-0.229$ & $0.004$ & $0.007$
& $-0.230\,(-0.375, -0.085)$
& $0.002$ & $0.003$ \\

F2 & Negative-AOI disengagement latency
& $-0.107$ & $0.187$ & $0.222$
& $0.037\,(-0.152, 0.226)$
& $0.701$ & $0.752$ \\

F2 & Late--early change in negative--positive gaze allocation bias
& $-0.109$ & $0.176$ & $0.221$
& $-0.106\,(-0.275, 0.064)$
& $0.222$ & $0.271$ \\

F2 & Late--early change in negative stickiness
& $0.029$ & $0.717$ & $0.752$
& $0.001\,(-0.159, 0.161)$
& $0.991$ & $0.991$ \\

F2 & High--low load change in negative--positive gaze allocation bias
& $0.018$ & $0.820$ & $0.820$
& $-0.011\,(-0.168, 0.146)$
& $0.890$ & $0.910$ \\

F2 & High--low load change in negative stickiness
& $-0.060$ & $0.461$ & $0.507$
& $-0.088\,(-0.258, 0.082)$
& $0.312$ & $0.371$ \\

% ===================== F3 =====================

F3 & Normalized scanpath length
& $0.213$ & $0.008$ & $0.012$
& $0.253\,(0.096, 0.411)$
& $0.002$ & $0.003$ \\

F3 & Normalized gaze convex-hull area
& $0.107$ & $0.183$ & $0.222$
& $0.121\,(-0.052, 0.293)$
& $0.170$ & $0.214$ \\

F3 & AOI transition duration
& $0.343$ & $<0.001$ & $<0.001$
& $0.344\,(0.183, 0.506)$
& $<0.001$ & $<0.001$ \\

F3 & Outside-AOI transition pause duration
& $0.312$ & $<0.001$ & $<0.001$
& $0.243\,(0.121, 0.364)$
& $<0.001$ & $<0.001$ \\

F3 & Within-AOI gaze step size
& $0.225$ & $0.005$ & $0.008$
& $0.242\,(0.090, 0.394)$
& $0.002$ & $0.003$ \\

F3 & Mean saccade amplitude
& $-0.303$ & $<0.001$ & $<0.001$
& $-0.302\,(-0.454, -0.150)$
& $<0.001$ & $<0.001$ \\

% ===================== R1 =====================

R1 & Target-hit proportion
& $-0.571$ & $<0.001$ & $<0.001$
& $-0.547\,(-0.671, -0.422)$
& $<0.001$ & $<0.001$ \\

R1 & Pre-target non-target AOI count
& $0.533$ & $<0.001$ & $<0.001$
& $0.423\,(0.299, 0.547)$
& $<0.001$ & $<0.001$ \\

R1 & Pre-target search path length
& $0.618$ & $<0.001$ & $<0.001$
& $0.485\,(0.339, 0.632)$
& $<0.001$ & $<0.001$ \\

% ===================== R2 =====================

R2 & Off-target dwell proportion
& $0.415$ & $<0.001$ & $<0.001$
& $0.382\,(0.252, 0.511)$
& $<0.001$ & $<0.001$ \\

R2 & Longest continuous target run
& $-0.525$ & $<0.001$ & $<0.001$
& $-0.495\,(-0.651, -0.339)$
& $<0.001$ & $<0.001$ \\

R2 & Target re-entry count after first hit
& $-0.031$ & $0.698$ & $0.749$
& $-0.033\,(-0.182, 0.116)$
& $0.665$ & $0.732$ \\

% ===================== R3 =====================

R3 & Strict target-fixation count
& $-0.185$ & $0.021$ & $0.032$
& $-0.169\,(-0.324, -0.014)$
& $0.033$ & $0.048$ \\

R3 & Total strict target-fixation duration
& $-0.661$ & $<0.001$ & $<0.001$
& $-0.622\,(-0.755, -0.489)$
& $<0.001$ & $<0.001$ \\

R3 & Candidate target-area fixation dispersion
& $0.273$ & $<0.001$ & $0.001$
& $0.280\,(0.130, 0.430)$
& $<0.001$ & $<0.001$ \\

R3 & Candidate target-area fixation step size
& $0.150$ & $0.063$ & $0.090$
& $0.173\,(0.014, 0.332)$
& $0.033$ & $0.048$ \\

\end{longtable}

\vspace{-5pt}
\noindent
\begin{minipage}{\textwidth}
\fontsize{7pt}{8pt}\selectfont
$\rho_s$ denotes Spearman correlation with continuous HAMD-17 total score.
Adjusted $\beta$ values were estimated using separate linear regression
models with standardized HAMD-17 total score as the predictor and each
standardized eye-movement feature as the outcome, adjusting for age, sex,
and years of education. Robust standard errors were used.
The $\beta$ coefficient represents the SD difference in the
eye-movement feature associated with a one-SD increase in HAMD-17
total score. FDR correction was applied separately to the Spearman
correlation tests and the adjusted regression tests across all 44
eye-movement features, including 19 primary and 25 supplementary
features. HAMD-17, the 17-item Hamilton Depression Rating Scale;
FDR, false discovery rate; CI, confidence interval.
\end{minipage}

\endgroup
\clearpage

\begingroup

% Compact formatting
\fontsize{7pt}{8pt}\selectfont
\setlength{\tabcolsep}{1.5pt}
\renewcommand{\arraystretch}{0.92}
\setlength{\extrarowheight}{0pt}

% Reduce whitespace around longtable
\setlength{\LTpre}{0pt}
\setlength{\LTpost}{0pt}

\begin{longtable}{@{}
    P{0.055\textwidth}
    P{0.255\textwidth}
    C{0.17\textwidth}
    C{0.055\textwidth}
    C{0.17\textwidth}
    C{0.055\textwidth}
    C{0.17\textwidth}
    C{0.055\textwidth}
@{}}

\caption{\textbf{Sensitivity analyses for primary eye-movement features.}}
\label{tab:s4_sensitivity}\\

\toprule
\makecell[l]{Features\\domain}
& Features
& \multicolumn{2}{c}{Unadjusted model}
& \multicolumn{2}{c}{HAMA-adjusted model}
& \multicolumn{2}{c}{Extreme-excluded} \\

\cmidrule(lr){3-4}
\cmidrule(lr){5-6}
\cmidrule(lr){7-8}

& & $\beta$ (95\% CI) & $q_{\mathrm{FDR}}$
& $\beta$ (95\% CI) & $q_{\mathrm{FDR}}$
& $\beta$ (95\% CI) & $q_{\mathrm{FDR}}$ \\

\midrule
\endfirsthead

\multicolumn{8}{@{}l}{%
\textbf{Supplementary Table \thetable\ (continued)}
}\\

\toprule
\makecell[l]{Features\\domain}
& Features
& \multicolumn{2}{c}{Unadjusted model}
& \multicolumn{2}{c}{HAMA-adjusted model}
& \multicolumn{2}{c}{Extreme-excluded} \\

\cmidrule(lr){3-4}
\cmidrule(lr){5-6}
\cmidrule(lr){7-8}

& & $\beta$ (95\% CI) & $q_{\mathrm{FDR}}$
& $\beta$ (95\% CI) & $q_{\mathrm{FDR}}$
& $\beta$ (95\% CI) & $q_{\mathrm{FDR}}$ \\

\midrule
\endhead

\midrule
\multicolumn{8}{r@{}}{%
\fontsize{7pt}{8pt}\selectfont Continued on next page
}\\
\endfoot

\bottomrule
\endlastfoot

% ===================== F1 =====================

F1 & Time to full AOI coverage
& $0.284\,(0.143, 0.425)$ & $<0.001$
& $0.265\,(0.084, 0.446)$ & $0.007$
& $0.302\,(0.159, 0.444)$ & $<0.001$ \\

F1 & First-pass AOI coverage ratio
& $-0.446\,(-0.596, -0.296)$ & $<0.001$
& $-0.530\,(-0.744, -0.316)$ & $<0.001$
& $-0.442\,(-0.603, -0.281)$ & $<0.001$ \\

F1 & AOI allocation entropy
& $-0.600\,(-0.732, -0.468)$ & $<0.001$
& $-0.747\,(-0.930, -0.565)$ & $<0.001$
& $-0.618\,(-0.750, -0.486)$ & $<0.001$ \\

% ===================== F2 =====================

F2 & Negative-positive gaze allocation bias
& $0.142\,(-0.029, 0.313)$ & $0.130$
& $0.140\,(-0.080, 0.359)$ & $0.269$
& $0.071\,(-0.070, 0.211)$ & $0.341$ \\

F2 & Negative AOI visit duration
& $-0.077\,(-0.237, 0.083)$ & $0.364$
& $-0.165\,(-0.368, 0.038)$ & $0.163$
& $-0.107\,(-0.273, 0.059)$ & $0.247$ \\

F2 & Negative stickiness after first hit
& $-0.008\,(-0.179, 0.162)$ & $0.923$
& $-0.055\,(-0.281, 0.171)$ & $0.668$
& $-0.049\,(-0.219, 0.122)$ & $0.575$ \\

% ===================== F3 =====================

F3 & Inter-AOI switch rate
& $-0.508\,(-0.653, -0.363)$ & $<0.001$
& $-0.545\,(-0.721, -0.369)$ & $<0.001$
& $-0.527\,(-0.674, -0.380)$ & $<0.001$ \\

F3 & AOI transition diversity
& $-0.548\,(-0.678, -0.418)$ & $<0.001$
& $-0.643\,(-0.829, -0.457)$ & $<0.001$
& $-0.560\,(-0.696, -0.425)$ & $<0.001$ \\

F3 & Transition tortuosity
& $0.315\,(0.186, 0.445)$ & $<0.001$
& $0.344\,(0.168, 0.521)$ & $<0.001$
& $0.366\,(0.209, 0.523)$ & $<0.001$ \\

F3 & Within-AOI gaze dispersion
& $0.158\,(0.014, 0.302)$ & $0.043$
& $0.132\,(-0.064, 0.328)$ & $0.253$
& $0.143\,(-0.005, 0.291)$ & $0.078$ \\

% ===================== R1 =====================

R1 & First target-entry latency
& $0.574\,(0.435, 0.713)$ & $<0.001$
& $0.628\,(0.477, 0.779)$ & $<0.001$
& $0.605\,(0.468, 0.742)$ & $<0.001$ \\

R1 & Initial target-entry accuracy
& $-0.599\,(-0.709, -0.490)$ & $<0.001$
& $-0.718\,(-0.863, -0.574)$ & $<0.001$
& $-0.600\,(-0.706, -0.493)$ & $<0.001$ \\

R1 & First target-fixation latency
& $0.594\,(0.461, 0.728)$ & $<0.001$
& $0.666\,(0.522, 0.810)$ & $<0.001$
& $0.609\,(0.487, 0.731)$ & $<0.001$ \\

% ===================== R2 =====================

R2 & Target dwell proportion
& $-0.699\,(-0.794, -0.604)$ & $<0.001$
& $-0.753\,(-0.884, -0.623)$ & $<0.001$
& $-0.678\,(-0.778, -0.578)$ & $<0.001$ \\

R2 & Target lock-on ratio
& $-0.474\,(-0.607, -0.340)$ & $<0.001$
& $-0.450\,(-0.615, -0.285)$ & $<0.001$
& $-0.409\,(-0.538, -0.280)$ & $<0.001$ \\

R2 & Post-hit off-target ratio
& $0.233\,(0.125, 0.340)$ & $<0.001$
& $0.245\,(0.029, 0.461)$ & $0.041$
& $0.257\,(0.108, 0.406)$ & $0.001$ \\

% ===================== R3 =====================

R3 & Target-area fixation dispersion
& $0.185\,(0.033, 0.337)$ & $0.025$
& $0.083\,(-0.121, 0.286)$ & $0.506$
& $0.176\,(0.015, 0.337)$ & $0.047$ \\

R3 & Within-target fixation step size
& $0.088\,(-0.065, 0.242)$ & $0.289$
& $0.029\,(-0.161, 0.219)$ & $0.767$
& $0.109\,(-0.052, 0.269)$ & $0.233$ \\

R3 & Target-area stability ellipse area
& $0.138\,(-0.032, 0.308)$ & $0.133$
& $0.082\,(-0.154, 0.318)$ & $0.555$
& $0.083\,(-0.075, 0.240)$ & $0.339$ \\

\end{longtable}

% ===================== Table notes =====================

\vspace{-5pt}
\noindent
\begin{minipage}{\textwidth}
\fontsize{7pt}{8pt}\selectfont
Sensitivity analyses were conducted for prespecified primary features.
Unadjusted models included ordered severity as the only predictor.
HAMA-adjusted models additionally included HAMA total scores alongside
age, gender, and years of education.
Extreme-excluded models repeated the primary analyses after excluding
extreme eye-movement values.
HAMA, Hamilton Anxiety Rating Scale;
FDR, false discovery rate;
CI, confidence interval.
\end{minipage}

\endgroup
\clearpage

\begingroup

% Compact formatting
\fontsize{7pt}{8pt}\selectfont
\setlength{\tabcolsep}{1.5pt}
\renewcommand{\arraystretch}{0.92}
\setlength{\extrarowheight}{0pt}

% Reduce whitespace around longtable
\setlength{\LTpre}{0pt}
\setlength{\LTpost}{0pt}

\begin{longtable}{@{}
    P{0.355\textwidth}
    C{0.105\textwidth}
    C{0.105\textwidth}
    C{0.105\textwidth}
    C{0.105\textwidth}
    C{0.085\textwidth}
    C{0.085\textwidth}
@{}}

\caption{\textbf{Recall-task performance across depression severity groups.}}
\label{tab:s5_recall_performance}\\

\toprule
& \shortstack{No\\($n=32$)}
& \shortstack{Mild\\($n=39$)}
& \shortstack{Moderate\\($n=49$)}
& \shortstack{Severe\\($n=35$)}
& $H$
& $p$ \\
\midrule
\endfirsthead

\multicolumn{7}{@{}l}{%
\textbf{Supplementary Table \thetable\ (continued)}
}\\

\toprule
& \shortstack{No\\($n=32$)}
& \shortstack{Mild\\($n=39$)}
& \shortstack{Moderate\\($n=49$)}
& \shortstack{Severe\\($n=35$)}
& $H$
& $p$ \\
\midrule
\endhead

\midrule
\multicolumn{7}{r@{}}{%
\fontsize{7pt}{8pt}\selectfont Continued on next page
}\\
\endfoot

\bottomrule
\endlastfoot

Correct recall trials out of 12, mean (SD)
& 11.09 (1.06)
& 8.95 (2.05)
& 6.98 (2.84)
& 5.03 (2.99)
& 71.97
& $<0.001$ \\

\end{longtable}

% ===================== Table notes =====================

\vspace{-5pt}
\noindent
\begin{minipage}{\textwidth}
\fontsize{7pt}{8pt}\selectfont
Correct recall trials were defined as trials in which the participant
oriented gaze to the cued target location during the recall phase.
Group comparisons were performed using the Kruskal--Wallis test.
$H$ denotes the Kruskal--Wallis test statistic.
\end{minipage}

\endgroup

\end{document}